\documentclass[aps,twocolumn,superscriptaddress,longbibliography,floatfix,nofootinbib]{revtex4-2}

\usepackage{nameref}
\usepackage[colorlinks=true,urlcolor=blue,citecolor=blue,allcolors=blue]{hyperref}

\usepackage{amsmath, mathtools, amsthm}
\usepackage{bbold}
\usepackage{esint}
\usepackage{amssymb}
\usepackage{braket}
\usepackage{bm}
\usepackage{graphicx}
\usepackage{tikz}
\usetikzlibrary{calc}
\usepackage{soul}
\usepackage[dvipsnames]{xcolor}
\usepackage{array}
\usepackage{slashed}
\usepackage{upgreek}
\usepackage{ulem}
\usepackage[noabbrev]{cleveref}
\usepackage{float}
\usepackage[nolong]{glossaries}
\usepackage{subcaption}
\usepackage{enumitem}
\usepackage{centernot}

\usetikzlibrary{positioning, shapes.geometric, arrows.meta,decorations.pathreplacing}
\usetikzlibrary{decorations.pathreplacing,decorations.markings}
\usetikzlibrary{arrows}
\usetikzlibrary{arrows.meta, shadows}
\glsdisablehyper 
\crefrangelabelformat{appendix}{#3#1#4--#5#2#6}

\renewcommand{\O}{\mathcal{O}}
\renewcommand{\L}{\mathcal{L}}

\renewcommand{\H}{\mathcal{H}}
\newcommand{\A}{\mathcal{A}}

\renewcommand{\P}{\mathcal{P}}
\renewcommand{\S}{\mathcal{S}}
\newcommand{\K}{\mathcal{K}}
\newcommand{\R}{\mathcal{R}}
\newcommand{\M}{\mathcal{M}}

\newcommand{\Z}{\mathbb{Z}}
\newcommand{\N}{\mathbb{N}}
\newcommand{\C}{\mathbb{C}}

\newacronym{LGT}{LGT}{\textit{lattice gauge theory}}
\newacronym{VQE}{VQE}{\textit{variational quantum eigensolver}}
\newacronym{NISQ}{NISQ}{\textit{noisy-intermediate scale quantum}}
\newacronym{HEA}{HEA}{\textit{hardware-efficient ansatz}}
\newacronym{QFT}{QFT}{\textit{quantum field theory}}
\newacronym{RG}{RG}{\textit{renormalization group}}
\newacronym{VAPOR}{VAPOR}{\textit{Variational Algorithm for Pauli Orbit Renormalization}}

\DeclareMathOperator{\Span}{span}

\DeclareMathOperator{\Supp}{Supp}
\DeclareMathOperator{\Orb}{Orbit}
\DeclareMathOperator{\id}{Id}
\DeclareMathOperator{\Proj}{Proj}

\theoremstyle{definition} 
\newtheorem{alg}{Algorithm}[section]

\begin{document}
\title{Implementing Hamiltonian Renormalization Group Flow on Quantum Computers with VAPOR}

\author{Federica Fragomeno}
\altaffiliation{These authors contributed equally to this work.}
\affiliation{Department of Physics, University of Alberta, Edmonton, Alberta T6G 2G1, Canada}

\author{Jorden Roberts}
\altaffiliation{These authors contributed equally to this work.}
\affiliation{Department of Physics, University of Alberta, Edmonton, Alberta T6G 2G1, Canada}

\author{Saeed Rastgoo}
\affiliation{Department of Physics, University of Alberta, Edmonton, Alberta T6G 2G1, Canada}
\affiliation{Department of Mathematical and Statistical Sciences, University of Alberta, Edmonton, Alberta T6G 2G1, Canada}
\affiliation {Theoretical Physics Institute, University of Alberta, Edmonton, Alberta T6G 2G1, Canada}

\author{Klaus Liegener}
\affiliation {Walther-Meissner-Institute, Walther-Meissner-Strasse 8, 85748 Garching, Germany}

\date{\today}

\begin{abstract}
While Hamiltonian Lattice Gauge Theory is gaining traction, today's limited numerical capacity leaves simulations affected by discretization errors. 
This motivates the implementation of renormalization group (RG) techniques to find discretization-error-free operators. To this end, we introduce VAPOR, a variational quantum algorithm that decomposes operators into Pauli strings, identifies RG flow orbits, and determines fixed points of a naively discretized operator. We illustrate this using a toy model of a kinematic operator in a symmetry-restricted SU(2) Yang-Mills theory.
\end{abstract}

\maketitle

\section{Introduction} 

One of the conceptual challenges in \gls{QFT} is the construction of interacting continuum theories. While perturbative expansions around the free vacuum have proven extremely successful, the mathematical existence of interacting quantum fields in four dimensions remains largely unresolved. This is prominently illustrated by the Yang–Mills mass gap problem \cite{JaffeWitten2000-YangMillsMassGap}, whose resolution depends on the representation of the Hamiltonian on a suitable continuum Hilbert space \cite{Haag1955,StreaterWightman2000}. However, most progress in simulating interacting \glspl{QFT} relies on discretizations, e.g., \gls{LGT} \cite{Wilson1974}, which regulate ultraviolet divergences. As naive discretizations of the Hamiltonian suffer from {\it discretization ambiguities}, any prediction must be validated against continuum physics\footnote{This is relevant where experiments are absent, e.g., different resolutions of the initial big-bang singularity in quantum gravity \cite{DaporLiegener2017,Ashtekar2006}.}. Despite remarkable numerical advances \cite{BrowerChrist2018LatticeQCDExascale,Rico2016,Yang2020,low2017optimal,Halimeh2025ColdAtomGauge,Halimeh2025OutOfEquilibriumGauge}, the computational complexity of simulating  strongly interacting systems remains computationally inaccessible beyond small lattices, making discretization effects difficult to control.

Conceptually, the relation between continuum and discrete theories is governed by Wilson's \gls{RG} program \cite{Wilson1975,Kadanoff1966}. Mathematically, this perspective relies on cylindrical consistency: given a family of Hilbert spaces and operators across scales, suitable consistency conditions allow the construction of a continuum theory via inductive limits \cite{Janas1988,KadisonRingrose1997,GlimmJaffe1987}. This sparks the question: if only a bare discretization is given, how can one change it to obtain a cylindrically consistent family of operators across scales? To solve this, Hamiltonian renormalization can be interpreted as a program for identifying operators stable under an \gls{RG} flow.

This differs from standard {\it coarse-graining} in condensed matter and high-energy physics \cite{Politzer1973,WilsonFisher1972}, where ``\gls{RG} flow'' often describes an operator's coarse-graining irrespective of cylindrical consistency: given a fundamentally discretized theory\footnote{Ising models \cite{Kadanoff1966,NiemeijerVanLeeuwen1973} or some discrete quantum gravity approaches \cite{Rastgoo:2016syw, Requardt:2015ila} feature \gls{RG} flows yielding effective theories mimicking continuum properties at large scales while being fundamentally discrete.} at a certain scale (or where discretization errors are deemed physically insignificant), \gls{RG} flow yields an effective theory for efficient numerical computations. However, lacking finer resolution data, this sub-family cannot reconstruct a continuum theory. Instead, this letter uses \gls{RG} methods to construct operator families admitting an inductive limit, ensuring their coarse-grained elements emerge from a continuum \gls{QFT}.

This application of \gls{RG} corresponds to the ``perfect action" in the covariant language \cite{HasenfratzNiedermayer1994,Polchinski1984}. Hamiltonian \gls{RG} formulations \cite{Hamiltonian_RG_I,Liegener21,Bahr_2022}, however, have received less attention because finding fixed points of interacting many-body Hamiltonians is computationally demanding. In this letter, we advocate that quantum computing provides new opportunities to address this problem. We formulate the fixed-point condition as a variational quantum optimization problem \cite{Peruzzo2014,McClean2016,Bharti2022} and demonstrate its feasibility in a toy model. This provides a route toward constructing cylindrically consistent Hamiltonians for the non-perturbative definition of continuum \gls{QFT}.

\section{Hamiltonian Renormalization\label{sec:RG}}
We introduce the basic concepts and notation of the Hamiltonian \gls{RG}. Readers familiar with the subject may jump ahead to \Cref{sec:alg}.

\begin{figure}[htp]
   \centering
    \definecolor{myblue}{RGB}{0,50,140}
    \definecolor{myred}{RGB}{123,12,22}
    \definecolor{myyellow}{RGB}{255,170,77}
    \definecolor{mygreen}{RGB}{33,105,33}
    \begin{tikzpicture}[scale=0.625, every node/.style={transform shape}]
        \filldraw[fill=myblue!5, draw=myblue, dashed, thick, rounded corners=3mm] 
            (8.75, 7.3) rectangle (12.05, -2)
            node[midway, above, yshift=4.65cm] {\textcolor{myblue}{\large\textbf{$\{O_M^*\}_{M\in 2^{\mathbb{N}_0}}$}}};
        \filldraw[fill=myred!15, draw=myred!15, thick, rounded corners=2mm] 
            (-0.5, 7) rectangle (0.35, -0.7) node[midway, above,xshift=0.25cm, yshift=3.85cm, myred, align=center, font=\bfseries] {\large\textbf{$\{O_M^{(0)}\}_{M\in 2^{\mathbb{N}_0}}$}};
        \node(A1) at (0,6.3) {\large $O_M^{(0)}$};
        \node(B1) at (0,4.8) {\large $O_{2M}^{(0)}$};
        \node(C1) at (0,1.5) {\large $O_{2^\ell M}^{(0)}$};
        \foreach \y in {3, -0.3}{
            \draw[line width=0.3mm, dotted] (0, \y) -- ++(90:0.4cm);
        }

        \foreach \x/\y/\i in {0.35/5.05/{2M\rightarrow M}, 2.6/5.05/{2M\rightarrow M}, 5/5.05/{2M\rightarrow M},
        0.35/3.6/{4M\rightarrow 2M}, 2.6/3.4/{4M\rightarrow 2M}, 5/3.35/{4M\rightarrow 2M},
        0.35/1.8/{2^\ell M\rightarrow 2^{\ell-1}M}, 2.6/1.78/{2^\ell M\rightarrow 2^{\ell-1}M}, 5/1.78/{2^\ell M\rightarrow 2^{\ell-1}M}}{
            \draw[-{Stealth[round,length=2.5mm]}, mygreen!85, thick] (\x, \y) --node[above,sloped]{$\mathcal{A}_{\i}$} ++(33:1.75cm); 
        }

        \begin{scope}[xshift=2.25cm]
            \node(A1) at (0,6.3) {\large $O_M^{(1)}$};
            \node(B1) at (0,4.8) {\large $O_{2M}^{(1)}$};
            \node(C1) at (0,1.5) {\large $O_{2^\ell M}^{(1)}$};
            \foreach \y in {3,-0.3}{
                \draw[line width=0.3mm, dotted] (0, \y) -- ++(90:0.4cm);
            }
            
        \end{scope}

        \begin{scope}[xshift=4.5cm]
            \foreach \y in {6.3, 4.8, 3, 1.5}{
                \draw[line width=0.3mm, dotted] (-0.1, \y) -- ++(0:0.4cm);
            }

        \end{scope}

        \begin{scope}[xshift=6.75cm]
            \node(A1) at (0,6.3) {\large $O_M^{(n)}$};
            \node(B1) at (0,4.8) {\large $O_{2M}^{(n)}$};
            \node(C1) at (0,1.5) {\large $O_{2^\ell M}^{(n)}$};
            \foreach \y in {3}{
                \draw[line width=0.3mm, dotted] (0, \y) -- ++(90:0.4cm);
            }
            \draw[line width=0.3mm, dotted] (0, -0.3) -- ++(90:0.4cm);   
            
        \end{scope}

        \begin{scope}[xshift=8cm]
            \foreach \y in {6.3, 4.8, 3, 1.5}{
                \draw[line width=0.3mm, dotted] (-0.1, \y) -- ++(0:0.4cm);
            }
        \end{scope}

        \begin{scope}[xshift=9.75cm]
             \node[circle, fill=white, inner sep=1pt, minimum size=7.5mm](A1) at (0,6.3) {\makebox[0pt]{\large $O_M^*$}};
            \node[circle, fill=white, inner sep=1pt, minimum size=7.5mm](B1) at (0,4.8) {\makebox[0pt]{\large $O_{2M}^*$}};
            \node[circle, fill=white, inner sep=1pt, minimum size=8.5mm](C1) at (0,1.5) {\makebox[0pt]{\large $O_{2^\ell M}^*$}};
            \foreach \y in {3, 0.5}{
                \draw[line width=0.3mm, dotted] (0, \y) -- ++(90:0.4cm);
            }

            \foreach \y/\i/\j in{5.2/{2M\rightarrow M}/-0.15cm, 3.6/{4M\rightarrow 2M}/-0.08cm,  1.95/{2^\ell M\rightarrow 2^{\ell-1}M}/+0.25cm}{
                \draw[-{Stealth[round,length=2.5mm]},  mygreen!85, thick] (0, \y) --node[align=center, xshift=1cm\j]{$\mathcal{A}_{\i}$} ++(90:0.7cm); 
            }
            
            \draw[-{Stealth[round,length=2.5mm]}, myblue, thick] (0, 0.25) --node[above, sloped, xshift=-0.15cm, yshift = 0.05cm]{$\ell\rightarrow\infty$} ++(-90:1.25cm);
            \node[circle, fill=myblue!8, inner sep=1pt, minimum size=8mm, draw=myblue](D4) at (0, -1.5) {\large \textcolor{myblue}{$O_\infty^*$}};
        \end{scope}
   \end{tikzpicture}
   \captionof{figure}{Operator families (columns) across different resolutions $M$ converge under the \gls{RG} flow $\mathcal{A}_{M\rightarrow M'}$ to a fixed-point family $\{O_M^*\}_M$ which allows a continuum limit $O_\infty^*$ (right).
   }
    \label{fig:RG}
\end{figure}
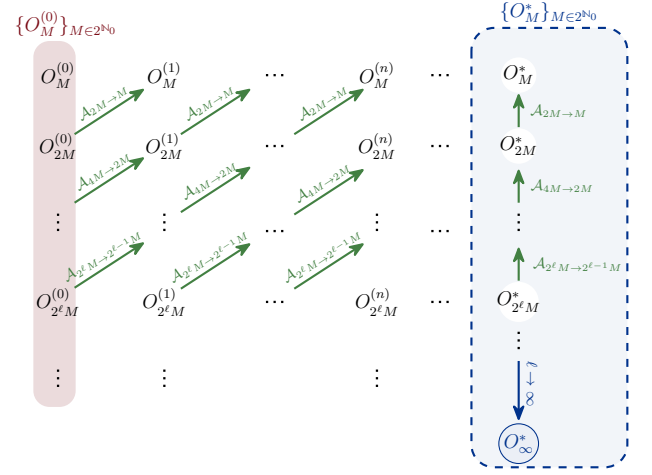

Constructive QFT \cite{Balaban1984I, Balaban1984II} aims at a rigorous representation of continuum observables $O_\infty$ on a ``continuum'' Hilbert space $\H_\infty$, featuring a vacuum state $\Omega_\infty$ annihilated by the Hamiltonian ($H_\infty\Omega_\infty=0$). We use the subscript ``$\infty$'' to label the infinitely fine resolution that these continuum objects carry. To find these representations, the \gls{RG} provides a pathway by first constructing families of {\it discretized} theories that allow continuum reconstruction via {\it inductive limits} \cite{Janas1988,KadisonRingrose1997,Hamiltonian_RG_I}: Assume a family of Hilbert spaces $\{\H_M\}_{M\in\mathcal{I}}$ indexed by finite resolution $M$ belonging to a partially ordered, directed set $\mathcal{I}$.
Different scales are related using isometric and cylindrically consistent \textit{embedding maps},
\begin{align}
    \label{embed}
    I_{M\to M'}:\H_M \to \H_{M'}, && M \leq M',
\end{align}
identifying coarse states with refined ones.
These maps define an \textit{inductive system} (see \Cref{app:RG}), $\H_\infty = \overline{\mathcal{D}}_\infty$, completing the pre-Hilbert space $\mathcal{D}_\infty \coloneqq \left(\coprod_{M\in\mathcal{I}} \H_M\right)/\sim$,
where states are identified via $\psi \sim I_{M \to M'}\psi$ \cite{Bahr_2022}. Similarly, continuum observables (such as the Hamiltonian) can be reconstructed from a family of {\it cylindrically consistent} operators $\{O_M\}_{M\in\mathcal{I}}$ that satisfy:
\begin{align}\label{cyl_cons_op}
    O_M = \A_{M' \to M}\left(O_{M'}\right) \coloneqq  I^\dagger_{M\to M'} O_{M'} I_{M\to M'},
\end{align}
where $\A_{M' \to M}$ is the {\it coarse-graining map} \cite{Hamiltonian_RG_I}.\footnote{In general, $O_\infty$ is only guaranteed to exist as a bilinear form, however in suitable cases the extension to an operator is possible. Here, we loosely refer to observables as operators and refer to \cite{Bahr_2022} for further details.}
Because families of operators built from naive discretizations generally fail to satisfy this consistency condition, they cannot naturally arise from an underlying continuum \gls{QFT}. To identify a cylindrically consistent family, the \gls{RG} program constructs one, starting from a naive initial discretization $O_M^{(0)}$, by iteratively applying the coarse-graining map, generating an {\it \gls{RG} flow} of the whole family
\begin{align} \label{op_flow}
    O^{(n+1)}_M \coloneqq \A_{M'\to M}\left(O^{(n)}_{M'}\right), && \forall n\in\N_0.
\end{align}
Ideally, this flow converges to operators $O^*_M \coloneqq \lim_{n\to\infty} O^{(n)}_M$ that remain invariant under the \gls{RG} map, satisfying
\begin{align} \label{fp_property}
    \A_{M' \to M}\left(O^*_{M'}\right) = O^*_M\,, && \forall M \leq M'.
\end{align}
Operators with this property are referred to as a \textit{fixed points} of the flow
. Provided that such fixed points exist, the resulting family $\{O^*_M\}_{M\in\mathcal{I}}$ is cylindrically consistent, thereby determining a continuum observable. See \Cref{fig:RG}.

However, implementing the Hamiltonian \gls{RG} faces three major practical challenges:
\setlist{nolistsep}
\begin{enumerate}[noitemsep]
    \item[\textbf{(1)}] The generation of increasingly non-local interactions during \gls{RG} flow necessitates a larger number of lattice vertices, drastically increasing the complexity of matrix element computations.
    \item[\textbf{(2)}] Starting from a naive discretization does not guarantee convergence to a physically relevant fixed point; some flows diverge or reduce to a free theory \cite{Aizenman1981,Landau1955}.
    \item[\textbf{(3)}] The dimension of the Hilbert space $\H_M$ grows rapidly with resolution $M$, making the explicit computation and storage of \gls{RG} flows computationally prohibitive.
\end{enumerate}
Overcoming these technical hurdles is necessary to fully leverage the Hamiltonian \gls{RG} program for defining continuum interacting \gls{QFT}.

\section{Variational Algorithm for Pauli Orbit Renormalization (VAPOR)\label{sec:alg}}
We now address the technical bottlenecks for operator-level construction of (cylindrically consistent) fixed points via \gls{RG} methods \cite{Hamiltonian_RG_I, Liegener21, Bahr_2022}. To handle the computations on large Hilbert spaces, we use quantum computers, benefiting from their ability to natively represent the \gls{QFT} Hilbert space on a sufficiently large qubit space \cite{Cerezo:2020jpv, Bharti2022, Rico2016, Cochran2025,joshi2026jad,xu2026jad}. Thus, we propose \gls{VAPOR} to make fixed-point identification computationally tractable.

To tailor everything to \gls{LGT}, we focus on locally-defined, translation-invariant operators $O_M$ on 
Hilbert spaces of the form $\H_M \cong \bigotimes_{m \in \Z_M} \H_{m}$ \cite{PhysRevD.11.395,Rothe:1992nt,PhysRevD.107.026014}:
\begin{align}
    \label{global_op}
    O_M = \sum_{m\in\Z_M}\left(\bigotimes_{r<m-\delta_0}\id_r\right)\otimes O_m \otimes \left(\bigotimes_{s > m + \delta_1}\id_s\right),
\end{align}
where $\delta_0,\delta_1 \in \Z_M$, $\id_m$ is the identity on $\H_m$, and $O_m$ acts on a local subspace $\H^\text{loc}_m \coloneqq \H_{m-\delta_0}\otimes\cdots \otimes \H_{m+\delta_1}$.
Assuming that local operators defined at distinct vertices are related by translations along the lattice, $O_M$ is determined by a single representative $O_m$. Thus, we analyze the \gls{RG} flow of $O_m$ in isolation, and reconstruct global quantities (e.g., fixed points $O_M^*$) as in \eqref{global_op} (see \Cref{app:RG}). We also assume a necessary finite local Hilbert space cutoff to enable numerical computations. While choice of basis and cutoff is problem-specific and will impact the final predictions, we keep our algorithm general to account for any such choice.

Under these assumptions, we can expand $O_m$ in a Pauli basis \cite{Peruzzo2014, McClean2016}, 
\begin{align}\label{local_op_decomp}
    O_m = \sum_\alpha c_\alpha\,P_\alpha,
\end{align}
where $\{P_\alpha\} \eqqcolon \P$ denotes a set of Pauli strings acting on $\H^\text{loc}_m$.
The \gls{RG} flow transforms this basis linearly \cite{Hamiltonian_RG_I,Bahr_2022}:
\begin{align}
    \label{pauli_RG_map}
    \A: P_\beta \mapsto \sum_\alpha A_{\alpha\beta} P_\alpha,
\end{align}
with coefficients $\{A_{\alpha\beta}\}$ determined by the chosen embedding $I_{M \to 2M}$ (see \Cref{app:alg}).

Viewing the Pauli coefficients $\{c_\alpha\}$ as a vector, the fixed-point condition \eqref{fp_property} reduces to an eigenvalue problem for the induced \gls{RG} matrix $A:\Span\P \to \Span\P$ \cite{Wilson1975}. To solve it, we use Krylov subspaces \cite{Krylov1931,Saad2011} to define ({\it forward}) {\it Pauli orbits}\footnote{Here we restrict to forward orbits generated by repeated application of the \gls{RG} map. In general, one may consider full orbits obtained by including inverse transformations.}, $\Orb_\A P_\alpha \subset \P$, for each Pauli string $P_\alpha \in \Supp O_m$ (see \Cref{app:alg_supp-orb}). Each orbit forms a finite-dimensional set of Pauli strings closed under $\A$, decoupling the fixed-point problem across orbits. For technical feasibility, computations might be restricted to order-$n_0$ Krylov subspaces by introducing a cutoff $n \leq n_0$ \cite{Krylov1931}.

Now, the Pauli orbits can be used to fully characterize the local \gls{RG} flow $\{O^{(n)}_m\}_{n\in\N_0}$. Specifically, we can write
\begin{align}
    \label{orbit-decomposition}
    O^{(n)}_m = \sum_{P_\alpha\in\Supp O_m} c_\alpha\,O_{\alpha}^{(n)}(\theta),
\end{align}
where each $O_{\alpha}^{(n)}(\theta) \coloneqq \sum_{P_i\in\Orb_{\A}P_\alpha}\left(A^n\theta\right)_i P_i$ is supported on $\Orb_{\A}P_\alpha$. Here, $\theta = \{\theta_i\}$ represents variational coefficients inside the orbit.

To identify fixed points within a given orbit, we define a residual operator:
\begin{align}
    \label{fp_res}
    \R(\theta) \coloneqq O_\alpha(\theta) - \A\big(O_\alpha(\theta)\big),
\end{align}
which measures the deviation from the fixed-point condition.
We then introduce the cost function
\begin{align} \label{cost_fcn}
    \mathcal{C}(\theta) \coloneqq \sum_I \big|\bra{\psi_I} \R(\theta) \ket{\psi_I}\big|^2,
\end{align}
using a set of reference states $\{\ket{\psi_I}\}$. 

By construction, $\mathcal{C}(\theta) \geq 0$, with equality holding \textit{iff} the residual $\R(\theta)$ annihilates all chosen reference states. An \textit{orbit-supported} fixed point is defined as $O^*_\alpha \coloneqq O_\alpha(\theta^*)$, where $\theta^*$ minimizes \eqref{cost_fcn}. This optimization enforces the fixed-point condition within the subspace determined by the reference states $\{\ket{\psi_I}\}$\footnote{Note that a fixed point satisfies $\bra{\psi}\R(\theta^*)\ket{\psi} = 0$ for \textit{all} states $\ket{\psi}$. In practice, however, it often suffices to evaluate the cost function on suitably chosen entangled states that probe the relevant operator structure \cite{Gard2019EfficientSS,Cerezo:2020jpv}.}. Since equation $\R(\theta^*) = 0$ is linear and homogeneous, the solution is unique up to an overall rescaling of the coefficients $\theta^*$, fixable by an appropriate convention (see \Cref{app:alg}).

Note that \gls{VAPOR} eases the variational cost by distributing the optimization parameters across independent Pauli orbits. Implementing the cost function $\mathcal{C}(\theta)$ on a quantum device via suitable measurements on prepared entangled states provides an efficient way to solve the fixed-point condition on increasingly large Hilbert spaces \cite{Bharti2022,Cerezo:2020jpv,Rico2016}.

\section{Renormalizing Non-Abelian LGTs via VAPOR\label{sec:test}}
To demonstrate \gls{VAPOR}'s capabilities, we test it in a simplified setting that nonetheless exhibits the complexities encountered when dealing with non-Abelian \gls{LGT}s \cite{PhysRev.96.191,Wilson1974,PhysRevD.11.395}. Specifically, upon enforcing the local Gauss constraint, the resulting gauge-invariant Hilbert space is spanned by coupled intertwiner states, preventing a straightforward tensor product decomposition into independent local Fock spaces \cite{Ashtekar_2021,Gambini:1996ik,Rov04}. Thus, reconstructing an inductive family of Hilbert spaces and renormalizing operators thereon becomes a highly non-trivial setting where \gls{VAPOR} can prove valuable. 

To investigate these complexities in a minimal setting, we focus on Yang-Mills theories with the gauge group $\mathrm{SU}(2)$ \cite{PhysRev.96.191,Rothe:1992nt}. The Kogut-Susskind Hamiltonian is defined on a $d\geq2$-dimensional lattice $\gamma_M$ with lattice spacing $1/M$ (see \Cref{app:LGT}) \cite{PhysRevD.11.395}:
\begin{equation}\label{eq:KS-Ham}
    H_M = g_E \sum_e H_E(e) - g_B \sum_{\Box}H_B(\Box) + \sum_m \Lambda^I(m) G_I(m)
\end{equation}
where the electric part $H_E$ acts on the lattice edges $e\in\gamma_M$, the magnetic part $H_B$ acts on the plaquettes (closed loops of four edges) $\Box$ of $\gamma_M$, and $G_I$ enforces Gauss’ law at each vertex $m\in\Z_M$. In the quantum theory, physical states are annihilated by this constraint, defining the gauge-invariant physical Hilbert space $\H_{G,M} \subset \H_M$. The kinematical Hilbert space is
\begin{align}
    \label{H_m_kin}
    \H_m \cong L_2\left(\mathrm{SU}(2)\times\mathrm{SU}(2), d\mu_H\otimes d\mu_H\right),
\end{align}
with $\mu_H$ denoting the Haar measure. Because the physical Hilbert space is infinite-dimensional, mapping the theory onto finite qubits requires a suitable choice of basis and truncation. Among existing proposals \cite{Andrea24,Zohar17,Fontana25}, we choose the irreducible representations (irreps) truncated at a maximum cutoff $j_\text{max}$ \cite{Gambini:1996ik,Rov04,Ashtekar_2004}. This makes the simulation computationally viable while preserving the dominant low-energy dynamics at strong couplings\footnote{While standard LGT often probes the continuum limit by simply tuning coupling constants (e.g., $g_E/g_B \ll 1$) without adding counterterms, ensuring the rigorous existence of an underlying continuum theory requires inclusion of the latter. Hence, the conventional parameter choices used in the literature do not directly translate to our framework.}.

\subsection{Toy model definition}
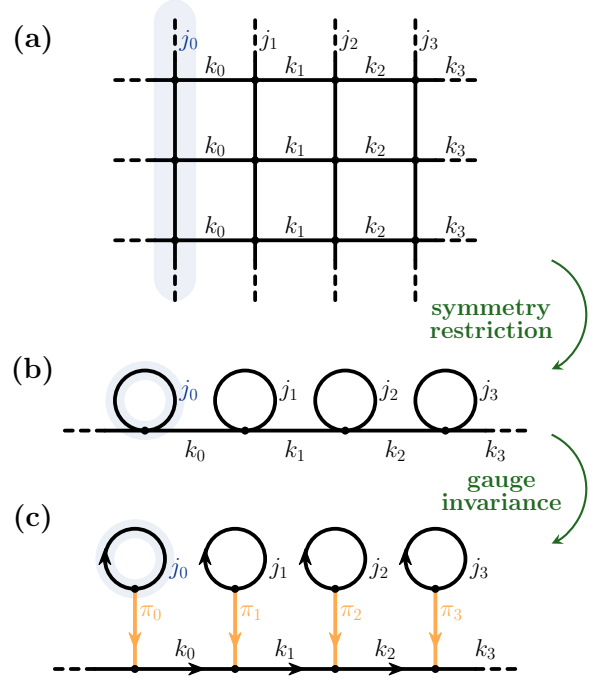
\begin{figure}[htp]
    \centering
    \definecolor{myblue}{RGB}{0,50,140}
    \definecolor{myred}{RGB}{123,12,22}
    \definecolor{myyellow}{RGB}{255,170,77}
    \definecolor{mygreen}{RGB}{33,105,33}
    \begin{tikzpicture}[scale=0.53, every node/.style={transform shape}, intertwiner/.style={line width=0.5mm, color=myyellow, line cap=round}]

        \filldraw[fill=myblue!7, draw=myblue!8, thick, rounded corners=3mm] 
            (-0.5, 4) rectangle (0.5, -3.5);
            
        \node at (-3.55, 3) {\huge\textbf{(a)}};
        \node at(0.35,3){\LARGE{\textcolor{myblue}{$j_0$}}};
        \node at(2.35,3){\LARGE{$j_1$}};
        \node at(4.35,3){\LARGE{$j_2$}};
        \node at(6.35,3){\LARGE{$j_3$}};

        \foreach \y in{-1.65, 0.35, 2.35}{
            \node at(1,\y){\LARGE{$k_0$}};
            \node at(3,\y){\LARGE{$k_1$}};
            \node at(5,\y){\LARGE{$k_2$}};
            \node at(7,\y){\LARGE{$k_3$}};
        }
        
        \foreach \y in{0,2, -2} {
            \draw[line width=0.5mm, line cap=round] (-0.5,\y) -- (6.5,\y);
            \draw[line width=0.5mm, dashed, line cap=round] (-0.5,\y) -- (-1.5,\y);
            \draw[line width=0.5mm, dashed, line cap=round] (6.5,\y) -- (7.5,\y);
        }
        \foreach \x in{0, 2, 4, 6} {
            \draw[line width=0.5mm] (\x,2.5) -- (\x,-2.5);
            \draw[line width=0.5mm, dashed, line cap=round] (\x,2.5) -- (\x,3.5);
            \draw[line width=0.5mm, dashed, line cap=round] (\x,-2.5) -- (\x,-3.5);
        }
        \foreach \x in {0, 4, 2, 6} {
            \foreach \y in {0,2, -2}{
                \node at (\x,\y) [circle,fill,inner sep=2pt]{};
            }
        }

        \draw[-stealth, thick, color=mygreen] (9, -2.75)++(0.4,0.25) arc[start angle=65, end angle=-65, radius=1.5] node[xshift=-1.5cm, yshift=1.25cm, align=center]{\textbf{\LARGE{symmetry}}\\\textbf{\LARGE{restriction}}};

        \begin{scope}[yshift=-6.75cm, xshift=-1.75cm]
        \node at (-1.8, 1.5) {\huge\textbf{(b)}};
            \filldraw[fill = white, draw=myblue!8, line width=2.5mm, rounded corners=3mm] (1, 0.75) circle (0.75);
            
            \node(A2) at(2.25,-0.5){\LARGE{$k_0$}};
            \node(A3) at (4.75,-0.5) {\LARGE{$k_1$}};
            \node(A4) at (7.25,-0.5) {\LARGE{$k_2$}};
            \node(A5) at (9.75,-0.5) {\LARGE{$k_3$}};
    
            \node(B1) at (2.1,1) {\LARGE{\textcolor{myblue}{$j_0$}}};
            \node(B2) at (4.6,1) {\LARGE{$j_1$}};
            \node(B3) at (7.1,1) {\LARGE{$j_2$}};
            \node(B4) at (9.6,1) {\LARGE{$j_3$}};

            \draw[line width=0.5mm, dashed, line cap=round] (-1,0) -- (0,0);
            \draw[line width=0.5mm, line cap=round] (0,0) -- (9.5,0);
            \draw[line width=0.5mm, dashed, line cap=round] (9.5,0) -- (10.5,0);
    
            \foreach \x in {1,3.5,6,8.5} {
                \draw[line width=0.5mm] (\x,0.75) circle (0.75); 
                \node at (\x,0) [circle,fill,inner sep=2pt]{};
            }
    
        \end{scope}

        \draw[-stealth, thick, color=mygreen] (9, -7.1)++(0.4,0.25) arc[start angle=65, end angle=-65, radius=1.5] node[xshift=-1.25cm, yshift=1.25cm, align=center]{\textbf{\LARGE{gauge}}\\\textbf{\LARGE{invariance}}};

        \begin{scope}[yshift=-12.7cm, xshift=-2cm]
            \node at (-1.55, 3.75) {\huge\textbf{(c)}};
            \filldraw[fill = white, draw=myblue!8, line width=2.5mm, rounded corners=3mm] (1, 2.75) circle (0.75);
            \node(A2) at(2.25,0.5){\LARGE{$k_0$}};
            \node(A3) at (4.75,0.5) {\LARGE{$k_1$}};
            \node(A4) at (7.25,0.5) {\LARGE{$k_2$}};
            \node(A5) at (9.75,0.5) {\LARGE{$k_3$}};

            \node(B1) at (2.1,2.5) {\LARGE{\textcolor{myblue}{$j_0$}}};
            \node(B2) at (4.6,2.5) {\LARGE{$j_1$}};
            \node(B3) at (7.1,2.5) {\LARGE{$j_2$}};
            \node(B4) at (9.6,2.5) {\LARGE{$j_3$}};

            \node(F1) at (1.4,1.4) {\textcolor{myyellow}{\LARGE{$\pi_0$}}};
            \node(F2) at (3.9,1.4) {\textcolor{myyellow}{\LARGE{$\pi_1$}}};
            \node(F3) at (6.4,1.4) {\textcolor{myyellow}{\LARGE{$\pi_2$}}};
            \node(F4) at (8.9,1.4) {\textcolor{myyellow}{\LARGE{$\pi_3$}}};

            \draw[line width=0.5mm, dashed, line cap=round] (-1,0) -- (0.0,0);
            \draw[line width=0.5mm, line cap=round] (0,0) -- (9.5,0);
            \draw[line width=0.5mm, dashed, line cap=round] (9.5,0) -- (10.5,0);
            \foreach \x in {2.25, 4.75, 7.25}{
                \draw[-{Stealth[round,length=2.5mm]}] (\x, 0) --  (\x+0.5,0);
            }
    
            \foreach \x in {1,3.5,6,8.5}{
                \draw[line width=0.5mm] (\x,2.75) circle (0.75); 
                \draw[-{Stealth[round,length=2.5mm]}] (\x-0.75, 3) --  ++(90:0.1mm);
                
                \draw[intertwiner] (\x,2) -- (\x,0);
                \draw[-{Stealth[round,length=2.5mm]}, myyellow] (\x, 1) --  (\x,0.5);
                \node at (\x,0) [circle,fill,inner sep=2pt]{};
                \node at (\x,2) [circle,fill,inner sep=2pt]{};
            }

        \end{scope}

    \end{tikzpicture}
    \caption{(a) Toy model based on a 2D lattice with vertical translational invariance: the irrep label $j_i$ applies across all vertical edges. (b) Symmetry restriction reduces the system to a 1D chain with one representative per vertical edge set. (c) The gauge-invariant basis is spanned by intertwiners ($\pi_i$, orange) coupling local irreps at each vertex.}
    \label{fig:lattice_reduction}
\end{figure}

To explore the \gls{RG} of non-Abelian gauge groups in higher dimensions, we construct a symmetry-restricted 2D Yang-Mills model. Restricting solutions on a two-dimensional spatial lattice to those with translational symmetry in the $y$-direction allows us to compactify this spatial direction by means of symmetry restriction \cite{Symmetry_Restriction,sph_sym_GoG}, reducing the system to a 1D chain (see \Cref{fig:lattice_reduction}). Here, vertical gauge links manifest as self-intersecting loops at each vertex, retaining the complexities of the 2D Gauss constraint and the non-trivial action of the magnetic loop within a manifestly 1D architecture. By inserting intertwiners at every node, we enforce gauge invariance \cite{Varshalovich1988,Brink:2023} and the gauge-invariant Hilbert space $\H_{G,M} \subset \H_M$ for any resolution $M\in\{2^\ell\}_{\ell\in\N_0}$ is spanned by states of the form
\begin{align}
    \ket{j, \pi, k}_M \coloneqq \ket{\{j_m, \pi_m, k_m\}_{m\in\Z_M}},
\end{align}
where each triple $(j_m, \pi_m, k_m)$ labels the representation data associated with a given vertex $m\in \Z_M$ (see \Cref{App:Toy_model}).
\hspace{-1em}

This model serves as a testbed for the \gls{RG} behaviour of the $\mathrm{SU}(2)$ intertwiner space and its operators. We focus on a translationally invariant family of local operators, each of which acts non-trivially only on a single vertex Hilbert space $\mathcal{H}_m$. The fundamental operators in $\mathrm{SU}(2)$ \gls{LGT} are the holonomy and flux operators \cite{Ashtekar_2004,Gambini:1996ik,Rov04}. For testing the \gls{RG} behaviour of gauge-invariant observables, we benefit from focusing on the latter. In particular, we construct a prototypical gauge-invariant and local operator: the \textit{4-flux operator}, $\O^{(0)}_{\rho, M}\in\L(\H_M)$. At each vertex $m\in\Z_M$, we set
\begin{align}
    \label{4-flux_loc}
    \O_{\rho,m}^{(0)} \coloneqq \iota_\rho^{(1111)}\cdot\bigotimes_{e\cap m \neq \emptyset} P(e),
\end{align}
where $\iota_\rho^{(j_1 j_2 j_3 j_4)}$ denotes a generic 4-valent intertwiner labelled by $\rho\in\N_0$ (see \Cref{app:LGT}), and $\cdot$ indicates complete contraction of magnetic indices \cite{Varshalovich1988,Balcar2009}. The global, translation-invariant $\O^{(0)}_{\rho,M}$ is obtained precisely as in \eqref{global_op}.

\subsection{Implementation of VAPOR on emulated QPU\label{sec:test_4-flux}}
We consider a particular embedding $I^\text{fill}_{M\to 2M}:\H_{G,M}\to\H_{G, 2M}$, called the \textit{filling kernel}, which acts on the gauge-invariant basis states \eqref{eq:state} according to
\begin{align}
    \label{local_embed}
    \ket{j,\pi,k}_M \longmapsto \ket{\{j_m,\pi_m,k_m,j_m,0,k_m\}_{m\in\Z_M}}.
\end{align}
This embedding preserves local connectivity and gauge invariance, facilitating a continuum Hilbert space description (see \Cref{app:RG}) \cite{Hamiltonian_RG_I,Liegener21}.

To characterize the full fixed-point family of 4-flux operators across all resolutions, one generally requires resolution-dependent Pauli coefficients. Since \eqref{4-flux_loc} is independent of $M$, however, the resolution dependence drops out in this toy model. Further, the analysis may be restricted to the minimal subgraph supporting the bare operator, namely a single vertex. Although the embedding $M \to 2M$ produces an intermediate operator on a larger lattice, translation invariance implies that it consists simply of two copies of the original operator located on the refined graph (see \Crefrange{app:RG}{app:alg}).

To determine the fixed point $\O^*_{\rho,m}$ with \gls{VAPOR} from \Cref{sec:alg}, we expand the bare operator in a local Pauli basis $\P$, as in \eqref{local_op_decomp}. By computing the orbits $\Orb_\A P_\alpha$ for each $P_\alpha \in \Supp \O^{(0)}_{\rho, m}$, we find $|\Orb_\A P_\alpha| \leq 2$
for any $P_\alpha\in\Supp \O^{(0)}_{\rho,m}$. 

Afterwards, we emulate a Quantum Processing Unit (QPU) via Qiskit's \textit{Statevector} simulator to evaluate the cost function \eqref{cost_fcn}, thereby identifying the fixed points independently within each orbit. We impose cutoffs of $j_\text{max}=1/2$ and $j_\text{max}=3/2$, which map the system to 6 and 12 qubits, respectively (see \Cref{app:alg_pauli-decomp}). To define the cost function, we use highly entangled vacua as reference states evaluated across several couplings $g_E=(1-\lambda^2)$ and $g_B=2\lambda^2$. The target ground states are then prepared using a parameterized \gls{HEA} optimized via the Powell algorithm, followed by a Trust Region Reflective (TRF) least-squares optimization of $\mathcal{C}(\theta)$ over each Pauli orbit (see \Crefrange{app:alg}{app:VQE}) \cite{Peruzzo2014,McClean2016,Cerezo:2020jpv,Kandala:2017vok}. 

As a benchmark for \gls{VAPOR}, we leverage our embedding choice to analytically compute the \gls{RG} flow of the 4-flux operator. The direct operator-level derivation of this fixed point is detailed in \Cref{app:RG}. \Cref{fig:paulicoeff} demonstrates that our algorithm successfully matches the analytical fixed points -- both recovering their structure when it differs from the bare operator and accurately identifying when the bare operator is already a fixed point (as in \Cref{fig:paulicoeff}\textcolor{blue}{b}).
\begin{widetext}
    \begin{minipage}{0.9\linewidth}
        \captionsetup{type=figure}
        \centering
        \includegraphics[width=\linewidth]{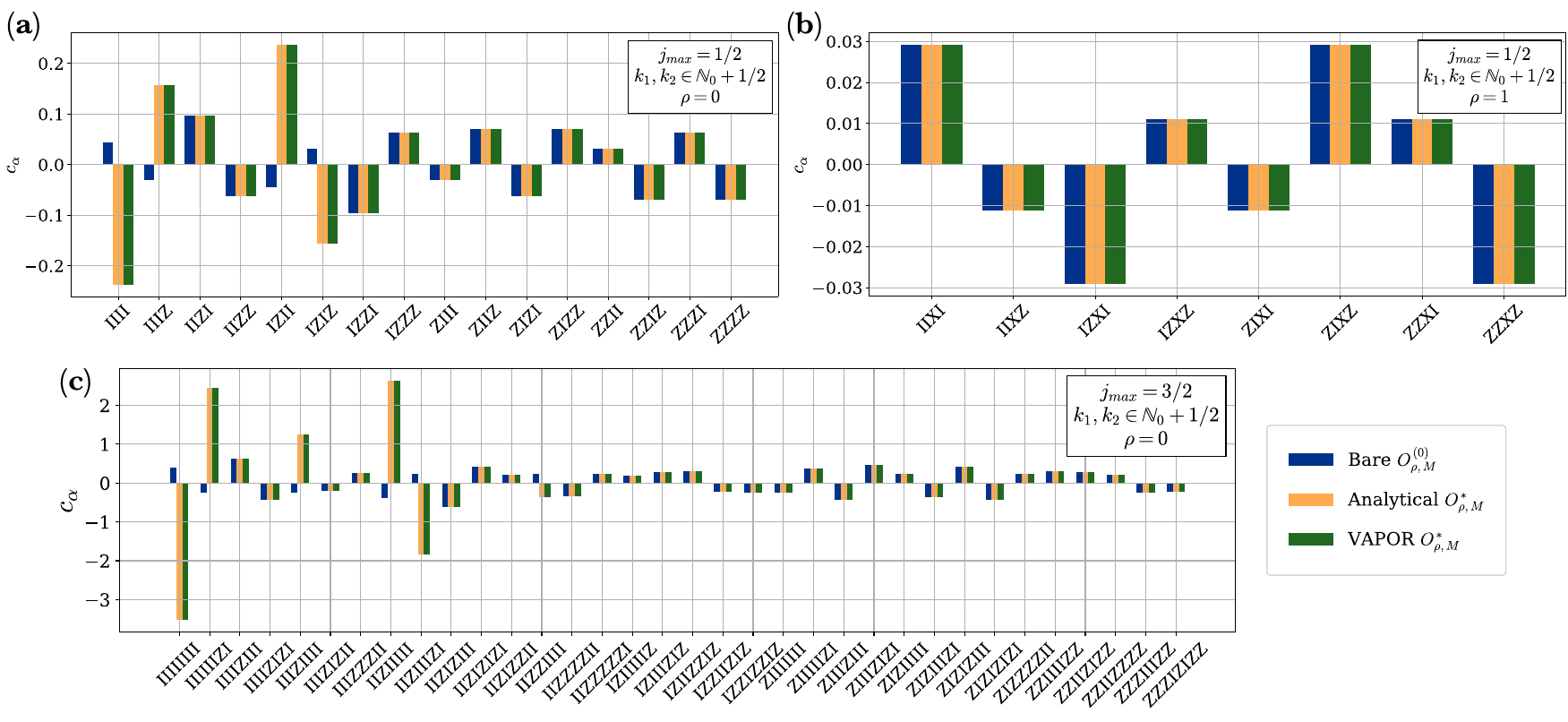}
        \captionof{figure}{Pauli coefficients $c_\alpha$ for the bare 4-flux operator (blue), its analytical fixed point (orange), and the \gls{VAPOR} fixed point (green). \gls{VAPOR} matches the analytical result exactly, regardless of operator type ($\rho=0,1$ in Figures \textcolor{blue}{3a}-\textcolor{blue}{3b}) and irrep cutoff (6 qubits for $j_\text{max}=1/2$ in Figures \textcolor{blue}{3a}-\textcolor{blue}{3b}; 12 qubits for $j_\text{max}=3/2$ in Figure \textcolor{blue}{3c}).}
        \label{fig:paulicoeff}
    \end{minipage}
\end{widetext}

\begin{figure}[htp]
    \centering
    \includegraphics[width=\linewidth]{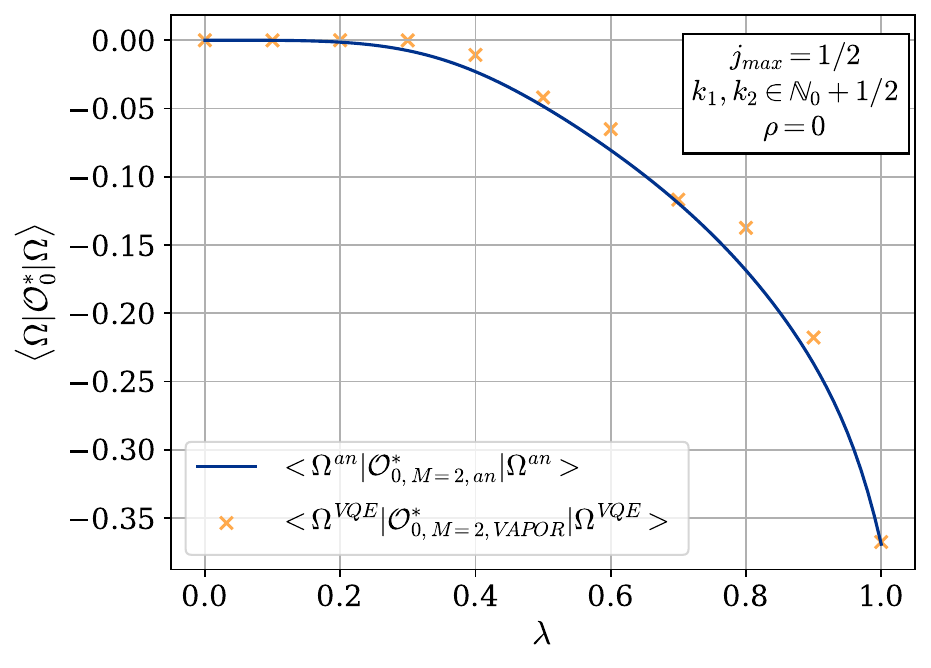}
   \caption{Fixed-point 4-flux operator $O^*_{0,M=2}$ as measurement for correlations on ground states. Analytical values (blue) are qualitatively matched by a noise-free QC-inspired simulation using \gls{VAPOR} and VQE (orange).
   }
    \label{fig:VQE_FP}
\end{figure}

This fixed-point operator, $\mathcal{O}^*$, can subsequently be used to probe system properties, such as correlations within the Kogut-Susskind vacuum. For this purpose, we employ a \gls{VQE} framework to prepare the target ground states $\ket{\Omega^{\text{VQE}}}$ variationally. By emulating a quantum device that measures $\mathcal{O}^*$ alongside variationally prepared \gls{LGT} ground states, we find good agreement between the expectation values of the \gls{VAPOR} operators computed on the VQE states, $\bra{\Omega^{\text{VQE}}} \mathcal{O}^*_{0,M=2,\text{VAPOR}}\ket{\Omega^{\text{VQE}}}$, and the exact analytical results, $\bra{\Omega^{\text{an}}} \mathcal{O}^*_{0,M=2,\text{an}}\ket{\Omega^{\text{an}}}$(see \Cref{fig:VQE_FP}). This level of consistency is supported by the high VQE state fidelity (see \Cref{app:VQE}). Ultimately, this demonstrates that physical continuum observables can be systematically extracted for near-term quantum simulations of \glspl{LGT}.

\section{Discussion}
To facilitate the construction of \gls{RG} fixed-point families -- and hence, the reconstruction of continuum \gls{QFT}s -- we introduced \gls{VAPOR}. It directly optimizes finite-resolution operators by decomposing them into Pauli strings and projecting them onto their eigenspaces, built as Krylov subspaces under the \gls{RG} map. Final eigenvectors are then determined variationally, where the computation of expectation values on large Hilbert spaces benefits significantly from quantum devices. This proposal was tested in a symmetry-restricted $\mathrm{SU}(2)$ Yang-Mills toy model. This model allowed the analytic renormalization of some gauge-invariant operators, such as the 4-flux operator, enabling a successful verification of the quantum algorithm's results before applying it to more complicated settings in forthcoming research.

We briefly discuss, on a qualitative level, why \gls{VAPOR} offers scaling advantages over classical numerics while simultaneously avoiding the challenges of comparable contemporary quantum algorithms:
\setlist{nolistsep}
\begin{enumerate}[noitemsep]
    \item[\textbf{(1)}] As the \gls{RG} flow of interacting \gls{LGT}s generates long-range interactions, describing operators requires enormous Hilbert spaces. Due to exponential scaling with increasing lattice size, classical numerical storage quickly becomes prohibitive. Representing observables and their \gls{RG} flow on quantum devices naturally circumvents these limitations.
    \item[\textbf{(2)}] Unlike other variational algorithms like \gls{VQE}, this proposal avoids the {\it barren plateau problem} \cite{McClean2018}, i.e. the exponential vanishing of derivatives when optimizing high-parameter cost functions. While VQE requires more parameters as the Hilbert space grows, \gls{VAPOR} directly ties the parameter count $n_0$ to the order-$n_0$ Krylov subspace built from the bare operator's relevant Pauli strings. As long as a small $n_0$ sufficiently approximates the fixed-point operator, barren plateaus are avoided\footnote{Note that identifying orbits depends on the choice of embedding map and basis. The filling kernel and Pauli mapping used in our toy model provide just one choice benefiting from a feasible orbit decomposition.}.
    \item[\textbf{(3)}] Like VQEs, \gls{VAPOR} utilizes shallow circuits since it only requires state preparation (benefiting from methods for highly entangled states, like $W$-states \cite{Romeiro2026ScalableWStates} or \gls{LGT} vacua \cite{Liegener:2026ryp}). In the noise-dominated era, this minimizes systematic errors (e.g., constant biases), as the algorithm probes the operator over a sufficiently large set $\{\ket{\Omega_I}\}$ without depending heavily on precise state preparation details.
\end{enumerate}

These features enable an efficient and scalable implementation of the \gls{RG} on near-term quantum hardware, providing a pragmatic route to operator-level renormalization in classically inaccessible regimes, including higher cutoffs and large lattice geometries.

Considering long-term prospects, in a fault-tolerant era, directly implementing embedding $I_{M\to2M}$ as a multi-qubit unitary will reduce the overhead of constructing Pauli orbits \textit{a priori}. Similarly, orbit optimization could be directly addressed by quantum singular value transformations \cite{gilyen2018qsvt,low2017optimal}. Given current hardware limitations, however, variational methods remain most suitable for exploring the \gls{RG} of Yang-Mills Hamiltonians, featuring long-range interactions and requiring larger qubit count, which is an application reserved for future research.

\section*{Acknowledgements}
We thank Jad Halimeh, Giovanni Cataldi, Philipp Aumann and Stefan Filipp for insightful discussions and helpful comments at various stages of this article. We are grateful for many discussions with the QC4HEP consortium. FF, JR and SR acknowledge the support of the Natural Sciences and Engineering Research Council of Canada (NSERC). KL acknowledges support by the Munich Quantum Valley, which is supported by the Bavarian state government with funds from the Hightech Agenda Bayern Plus.

\bibliography{mainbib.bib}
\appendix
\glsresetall 

\section{The Lattice Gauge Theory Framework\label{app:LGT}}
\subsection{Hamiltonian Formulation and Spin-Network Basis}
To construct a quantum simulation, we rely on the Hamiltonian formulation of LGT, originally introduced by Kogut and Susskind \cite{PhysRevD.11.395}. We consider a non-Abelian $\mathrm{SU}(2)$ gauge theory, which is distinguished by the non-commutativity of its local symmetry generators \cite{Peskin:1995ev},
\begin{equation}
[\tau^I, \tau^J] = i f^{IJ}_{\phantom{IJ}K} \tau^K
\end{equation}
with $I,J,K$ internal $\mathrm{SU}(2)$ indices. Here, $\tau^I \in \mathfrak{su}(2)$ are the Hermitian generators of the gauge group, and $f^{IJK}$ are its completely antisymmetric structure constants. For $\mathrm{SU}(2)$, the generators are proportional to the Pauli matrices ($\tau^I = -\sigma^I/2$), and the structure constants correspond to the Levi-Civita tensor, $f^{IJK} = \epsilon^{IJK}$. Crucially, this non-commuting algebraic structure manifests dynamically as self-interactions among the gauge fields. The non-Abelian field strength tensor necessarily acquires a non-linear term proportional to the coupling constant $g$,
\begin{equation}
F_{\mu\nu}^I = \partial_\mu A_\nu^I - \partial_\nu A_\mu^I + g\, \epsilon^I_{\phantom{I}JK} A_\mu^J A_\nu^K
\end{equation}
which fundamentally distinguishes the dynamics from those of free Abelian electrodynamics. Performing a  Legendre transformation on the Yang-Mills Lagrangian density $\L = -\frac{1}{4} F_{\mu\nu}^I F^{\mu\nu}_I$ \cite{PhysRev.96.191}, the continuum Hamiltonian for this $\mathrm{SU}(2)$ theory takes the form
\begin{equation}
    \begin{split}
        &H := \int d^3x (E_I^\mu(x)\dot{A}_\mu^I(x) - \L(x)) \\
        &=\int d^3x \left(\frac{1}{2}E_I^aE_a^I + \frac{1}{4} F_I^{ab}F_{ab}^I - A_0^I(\partial_aE_I^a - g\,\epsilon_{IJ}^{\phantom{IJ}K}A_a^JE_K^a)\right)
    \end{split}
    \label{eq:app-ham}
\end{equation}
where the indices $a,b$ denote spatial coordinates.

The $\mathrm{SU}(2)$ Hamiltonian is partitioned into two terms. The first, analogous to the continuous electric field energy, is the electric part, $H_{el}$, the second, corresponding to the field strength tensor, is the magnetic part, $H_{mag}$.
We discretize the spatial continuum onto a lattice and quantize the theory. Following Kogut-Susskind, the continuous gauge fields are mapped to group elements, holonomies $h(e)$, residing on the edges ee
e of the lattice. Conversely, the electric fields are mapped to their conjugate momenta, the electric gauge-covariant fluxes $P_I(e)$ \cite{Bojowald_2010, Thiemann_MCQGR, QSD_VII}. This discretization yields the Hamiltonian
\begin{equation}
    H = \frac{Mg^2}{2} \left(\sum_e P_I(e)P^I(e) + \frac{4}{g^4} \sum_\Box (2-tr[h(\Box)])\right)
\end{equation}
where the first sum runs over all lattice edges $e$, and the second sum runs over all elementary plaquettes. Here, $M$ is the lattice resolution parameter and $h(\Box)$ represents the plaquette holonomy.

In lattice theory, an element $g_e \in \mathrm{SU}(2)$ is associated with each lattice edge $e$. The state of a single edge is thus described by a wave function $\psi(g_e)$ residing in the single-vertex Hilbert space \eqref{H_m_kin}. By the Peter-Weyl theorem \cite{peter-weyl}, this space of square-integrable functions can be parameterized using a complete orthogonal basis of Wigner $D$-functions, $D^{(j)}_{pq}(g_e)$. These are labelled by the $\mathrm{SU}(2)$ irreducible representations (irreps) of discrete half-integer spin $j \in \{0, 1/2, 1, \dots\}$ and their corresponding magnetic quantum numbers $p$ and $q$ \cite{Brink:2023, Varshalovich1988}. Consequently, any generic edge function can be expanded as
\begin{equation}
    \psi(g_e) = \sum_{j,p,q} c^{(j)}_{pq}D^{(j)}_{pq}(g_e)
\end{equation}
Having parameterized the single-edge Hilbert space, we proceed to quantize the local degrees of freedom by promoting the classical electric and magnetic variables to quantum operators. The holonomy operator $h(e)$ acts as a multiplication operator, operating on the state by multiplying the wave function by the matrix elements of the $\mathrm{SU}(2)$ representation
\begin{equation}
    h(e)\psi(g_e) = D^{(j)}(g_e)\psi(g_e)
\end{equation}
Conversely, the electric flux operators $P^I(e)$ act as derivative operators on the group manifold, analogous to quantum angular momentum. Their action on a Wigner $D$-matrix is governed by the insertion of the Lie algebra generators $\tau^I$ as follows
\begin{equation}\label{flux_action}
    P^I(e)D^{(j)}_{pq}(g_e) = D^{(j)}_{ps}(g_e)\pi^{(j)}_{sq}(\tau_I)
\end{equation}
where $\pi^{(j)}(\tau^I)$ denotes the irreducible representation of the algebra $\mathfrak{su}(2)$ corresponding to the spin $j$. Thus, applying the squared electric flux yields
\begin{equation}
    \begin{split}
        P_I(e)P^I(e)D^{(j)}_{pq}(g_e) &= D^{(j)}_{ps}(g_e)\pi^{(j)}_{st}(\tau_I)\pi^{(j)}_{tq}(\tau_I) \\
        &= D^{(j)}_{ps}(g_e)\pi^{(j)}_{sq}(\tau_I\tau^I) \\
        &= j(j+1)D^{(j)}_{pq}(g_e)
    \end{split}
\end{equation}
Consequently, the electric part of the Hamiltonian is strictly diagonal in the spin-network basis, yielding an energy contribution proportional to a factor of $j(j+1)$ for the electric field on each link.

Evaluating the magnetic part is considerably more involved, as it introduces off-diagonal transitions. The relevant plaquette operators are defined by the traces of the holonomies around the closed loops
\begin{align}
    H_{\square,M} &\coloneqq \frac{1}{M}\sum_{m\in\Z_M}\mathrm{Tr}\Big[D^{(1/2)\dagger}(g_m) D^{(1/2)\dagger}(h_{m+1}) \nonumber \\
    & \qquad\qquad \times D^{(1/2)}(g_m) D^{(1/2)}(h_m)\Big].
    \label{plaquette}
\end{align}
where $g_m$ denotes the group element on the transverse direction and $h_m$ the one on the orthogonal one.

For a full lattice, the total unconstrained Hilbert space is the tensor product of the individual spaces, $\H_M \cong \bigotimes_{m\in \Z_M} \H_m$. However, this unconstrained space contains many redundant configurations that violate Gauss's law, meaning they are not gauge-invariant at the vertices. In a gauge theory, physical states must be invariant under local $\mathrm{SU}(2)$ transformation. This is enforced at each vertex $m$ by the Gauss constraint: 
\begin{equation}
    G_m \, |\psi\rangle = 0
\end{equation}
where $G_m$ is the last term of Eq.~\eqref{eq:app-ham}.

To systematically restrict ourselves to this physical subspace, we use the spin-network representation. In this framework, the basis is constructed by assigning each edge $e$ a specific irrep $j_e$, which acts on a $(2j_e+1)$-dimensional vector space $V_{j_e}$. The Gauss constraint then dictates that the angular momenta of all edges incident at a vertex $m$ must couple together to form the trivial representation (the $j=0$ singlet). This exact coupling is enforced by introducing an intertwiner $\iota_{\pi_m}$ at each vertex -- an invariant tensor belonging to the local intertwiner space $\mathcal{I}_m$ defined as
\begin{equation}
    \iota_{\pi_m} \in \mathcal{I}_m \equiv \text{Inv}_{\mathrm{SU}(2)}\left(\bigotimes_{e\owns m} V_{j_e}\right)
    \label{eq:intertw}
\end{equation}
where $\text{Inv}(\cdot)$ denotes the invariant subspace under $\mathrm{SU}(2)$.  
Assigning an intertwiner to each vertex projects the tensor product of edge representation spaces onto its gauge-invariant component, thereby reducing the unconstrained Hilbert space to the physical, gauge-invariant subspace
\begin{equation}
\begin{split}
    \mathcal{H}_{\text{phys}} =& \Span\left\{ \left( \bigotimes_{m} \iota_{\pi_m} \right) \cdot \left( \bigotimes_{e} D^{(j_e)}(g_e) \right) \,\right\}
\end{split}
\label{eq:H-phys}
\end{equation}
where dot $(\cdot)$ denotes the contraction of the group indices of the Wigner $D$-functions by the vertex intertwiners. 

The Kogut-Susskind Hamiltonian is inherently gauge-invariant, so its action is confined within this physical Hilbert space. Constructing it without intertwiners would force us to represent the full, unconstrained Hilbert space, bloating the matrix dimension with decoupled unphysical sectors and increasing the computational overhead. By building our basis states with intertwiners (known as spin-network basis encoding the quantum degrees of freedom) we guarantee gauge invariance by construction. To make the notation explicit, we label gauge-invariant basis states by their spin-network quantum numbers. An arbitrary vertex basis state $\ket{q}_m$ is then specified by assigning an irrep and an intertwiner:
\begin{equation}
    \ket{q}_m \coloneqq \ket{\{j_m\}, \{\pi_m\}}
\end{equation}
In this notation, $\{j_m\}$ denotes the collection of representation labels for the specific edges attributed to vertex $m$, rather than global edge indices. Furthermore, $\{\pi_m\}$ labels the choice of invariant intertwiner within the space defined by Eq.~\eqref{eq:intertw}. By transitioning to this vertex notation, the intertwiner labels $\pi_m$ distinguish an orthogonal basis within the invariant subspace at each respective vertex, streamlining the description of the local gauge-invariant degrees of freedom.

\subsection{Toy model}
\label{App:Toy_model}
To construct a computationally tractable model for quantum simulation while retaining the core $\mathrm{SU}(2)$ dynamics, we reduce the general formalism to a simplified geometry exhibiting homogeneity along one spatial direction. To this end, let $\gamma_M$ denote a 2D cubic lattice with periodic vertex set $\Z_M\times\Z_M$. At the classical level, discretizing an $\mathrm{SU}(2)$ Yang-Mills theory over $\gamma_M$ yields a symplectic manifold $(\M_M,\omega_M)$, with phase space $\M_M$ and symplectic form $\omega_M$ \cite{Ashtekar_2004,Thiemann_MCQGR,sph_sym_GoG}. In particular, we have 
\begin{align}
    \M_M \cong \prod_{m\in\Z_M^2}\big(\M_{m,x}\times\M_{m,y}\big),
\end{align}
and
\begin{align}
    \omega_M = \sum_{m\in\Z_M^2}\left(\omega_{m,x} + \omega_{m,y}\right),
\end{align}
where $\M_{m,i}\cong T^* SU(2)$ is the local phase space associated with the edge at vertex $m = (m_x,m_y)\in\Z_M^2$ oriented along the $i$-direction, which carries the canonical symplectic form $\omega_{m,i}$ \cite{QSD_VII}.

Following the framework of \textit{symmetry restriction} \cite{Symmetry_Restriction}, we impose translational symmetry along the spatial $y$-axis. More precisely, we restrict $(\M_M,\omega_M)$ to the invariant submanifold $(\overline{\M}_M,\omega_M|_{\overline{\M}_M})$, where $\overline{\M}_M \subset \M_M$ consists only of configurations invariant with respect to a symplectic analog of the discrete $y$-translation group $\mathbb{T}_M^y \cong \Z_M$ \cite{Symmetry_Restriction,sph_sym_GoG}. Such configurations are independent of the $m_y$-coordinate, so that $\omega_M|_{\overline{\M}_M}$ reduces to a sum over $m_x\in\Z_M$ alone, and we obtain
\begin{align}
    \label{M_bar}
    \overline{\M}_M \cong \prod_{m_x\in\Z_M}\left(\overline{\M}_{m_x,x}\times\overline{\M}_{m_x,y}\right).
\end{align}
Notably, all $y$-edges at fixed $m_x$ are now collectively represented by a single point in $\overline{\M}_{m_x,y}$. Imposing partial homogeneity therefore causes the family of edges along the symmetry direction to collapse to a single loop at each $m_x$ (see \Cref{fig:lattice_reduction}).

Because the Hamiltonian \eqref{eq:KS-Ham} is translation-invariant, the restricted system $(\overline{\M}_M, \omega_M|_{\overline{\M}_M},H_M|_{\overline{\M}_M})$ forms a consistent Hamiltonian subsystem of $(\M_M,\omega_M,H_M)$ \cite{Symmetry_Restriction}. Initially symmetric configurations are thus guaranteed to remain symmetric throughout their evolution, and the resulting dynamics can be described entirely within the reduced system. Geometrically, we have therefore reduced $\gamma_M$ to a 1-dimensional chain containing a self-loop at each vertex. This reduction drastically decreases the number of independent degrees of freedom while preserving the non-Abelian structure of the underlying LGT. In what follows, we assume that the reduced system continues to provide a valid effective description after quantization. Moreover, we drop the directional subscript and simply label vertices in the 1D graph by $m\in\Z_M$.

In this compactified geometry, the generic lattice edges $e$ take on two distinct physical roles. The transverse gauge fields, wrapped by the periodic boundary conditions, manifest as self-intersecting loops at each local vertex $m$, labelled by the irrep $j_m$. Conversely, the longitudinal spatial dimension forms the horizontal links connecting adjacent vertices, labelled by $k_m$. This explicit reassignment of the general edge degrees of freedom, $\{j_{e_x}, j_{e_y}\} \to \{j_m, k_m\}$, is illustrated in Figure \Cref{fig:lattice_reduction}\textcolor{blue}{b}.

Consequently, the total kinematical (unconstrained) Hilbert space, originally defined as the tensor product over all generic vertex $m$, can now be explicitly written in terms of this 1D geometry:
\begin{equation}
\H_{\text{kin}} = \bigotimes_m \H_m = \bigotimes_m \H_{j} \otimes \H_k
\end{equation}
where each local Hilbert space $\H_{j}$ and $\H_{k}$ carries the irreducible representations of $\mathrm{SU}(2)$ corresponding to the loop and edge momenta at vertex $m$, respectively. Then, imposing the Gauss’ law at each vertex yields the physical Hilbert space \eqref{eq:H-phys}.

The fully gauge-invariant quantum state for this 1D chain can be analytically expressed as a tensor contraction over all magnetic indices. Imposing boundary conditions to obtain a closed chain of $m$ vertices, the state of the full system reads
\begin{align}
    \ket{j,\pi,k}_M &\coloneqq  \sum_r\Bigg[\prod_{m\in\Z_M} \sqrt{d_{j_m}d_{\pi_m}d_{k_m}} \nonumber \\
    & \times \left(\psi_{\pi_m}^{(k_{m-1} j_m k_m)}(h_m,g_m)\right)_{r_{m-1} r_m}\Bigg],
    \label{eq:state}
\end{align}
with 
\begin{align} 
    \label{psi_vertex}
    &\left(\psi_{\pi_m}^{(k_{m-1} j_m k_m)}(h_m,g_m)\right)_{r_{m-1}r_m} \coloneqq \sum_{p,q,s}\sum_{q',s'}\left(\begin{array}{c}
         j_m  \\
         q'\;q 
    \end{array}\right) \nonumber \\
    &\qquad \times \left(\begin{array}{c}
         k_m  \\
         s'\;s 
    \end{array}\right)\left(\iota_{\pi_m}^{(j_m j_m k_{m-1} k_m)}\right)_{p q' r_{m-1} s'} \nonumber \\
    &\qquad\qquad \times D^{(j_m)}_{p q}(h_m) D^{(k_m)}_{r_m s}(g_m),
\end{align}
and 
\begin{align}
    \left(\iota_\pi^{(j_1 j_2 j_3 j_4)}\right)_{n_1 n_2 n_3 n_4} &\coloneqq \sum_{p,q}\left(\begin{array}{ccc}
        j_1 & j_2 & \pi \\
        n_1 & n_2 & p
    \end{array}\right)\left(\begin{array}{c}
         \pi  \\
         p \; q 
    \end{array}\right) \nonumber \\ 
    &\qquad\qquad \times \left(\begin{array}{ccc}
        \pi & j_3 & j_4 \\
        q & n_3 & n_4
    \end{array}\right).
\end{align}
In this expression, the three-column arrays represent the Wigner $3j$-symbols \cite{wigner1965group, Varshalovich1988, Brink:2023}, which define the components of the intertwiners. The single-column arrays denote the $\mathrm{SU}(2)$ metric tensors (or $1j$-symbols), which handle the relative orientations of the links. 

With the state formally defined, we proceed to compute the Hamiltonian matrix elements. The electric part of the Hamiltonian is straightforward to evaluate. It acts diagonally on the spin-network basis, yielding the sum of the quadratic Casimir eigenvalues for each respective physical edge
\begin{align}
    &{}_M\!\bra{j', \pi', k'} H_\text{el}(\gamma_M)\ket{j,\pi,k}_M = \frac{1}{M}\left(\prod_{i\in\Z_M}\delta_{j_i j_i'}\delta_{k_i k_i'}\delta_{\pi_i \pi_i'}\right) \nonumber \\
    &\qquad\qquad \times \sum_{m\in\Z_M}\big[j_m (j_m + 1) + k_m (k_m + 1)\big].
    \label{eq:H-el}
\end{align}
Turning to the magnetic Hamiltonian, we evaluate the action of the plaquette operator defined in Eqs. \eqref{plaquette} on our basis state \cite{Brink:2023, Bambi:2023jiz, Balcar2009}. By subsequently projecting the result onto the final state, we obtain the following matrix elements
\begin{align}
    &{}_M\!\bra{j',\pi',k'}H_\square(\gamma_M)\ket{j,\pi,k}_M = \frac{1}{M}\sum_{m\in\Z_M}\Bigg(\prod_{p\neq m}\delta_{k_p k_p'}\Bigg) \nonumber \\
    &\times \Bigg(\prod_{q\neq m,m+1}\delta_{j_q j_q'}\delta_{\pi_q \pi_q'}\Bigg)(-1)^{k_m' - k_m}(-1)^{\pi_m + \pi_{m+1}'} \nonumber \\
    &\times \sqrt{d_\square^{(m)} {d'}_\square^{(m)}} \sum_{\alpha,\beta}(-1)^{\alpha-\beta}d_\alpha d_\beta\sum_u d_u \left\{\begin{array}{ccc}
        j_m & j_m & \pi_m \\
        1/2 & \alpha & j_m'
    \end{array}\right\} \nonumber \\
    & \times \left\{\begin{array}{ccc}
        j_m' & j_m' & \pi_m' \\
        1/2 & \alpha & j_m
    \end{array}\right\}\left\{\begin{array}{ccc}
        k_{m-1} & k_m & \pi_m \\
        1/2 & \alpha & u
    \end{array}\right\}\left\{\begin{array}{ccc}
        k_{m-1}' & k_m' & \pi_m' \\
        1/2 & \alpha & u
    \end{array}\right\} \nonumber \\
    &\times \left\{\begin{array}{ccc}
        k_{m+1} & k_m & \pi_{m+1} \\
        1/2 & \beta & u
    \end{array}\right\}\left\{\begin{array}{ccc}
        k_{m+1}' & k_m' & \pi_{m+1}' \\
        1/2 & \beta & u
    \end{array}\right\} \nonumber \\
    &\times \left\{\begin{array}{ccc}
        j_{m+1} & j_{m+1} & \pi_{m+1} \\
        1/2 & \beta & j_{m+1}'
    \end{array}\right\}\left\{\begin{array}{ccc}
        j_{m+1}' & j_{m+1}' & \pi_{m+1}' \\
        1/2 & \beta & j_{m+1}
    \end{array}\right\},
    \label{eq:H-mag}
\end{align}
where $d_\square^{(m)} = d_{j_m}d_{\pi_m}d_{k_m}d_{j_{m+1}}d_{\pi_{m+1}}$ and ${d'}_\square^{(m)} = d_{j_m'}d_{\pi_m'}d_{k_m'}d_{j_{m+1}'}d_{\pi_{m+1}'}$ and the sums run over the internal angular momentum variables $u$, $\alpha$, and $\beta$ introduced during the re-coupling sequence. The quantities denoted by the curly brackets are the $6j$-symbols.

To make the quantum simulation computationally feasible, we further exploit the translational invariance of the lattice in \Cref{fig:lattice_reduction}. This allows us to restrict to the minimal non-trivial repeating unit of the system, i.e. two vertices. We  close the system by imposing periodic boundary conditions. In this minimal setup, the analytical expression for the initial state, as well as the electric and magnetic parts of the Hamiltonian, follow directly from evaluating the generalized Eqs.~\eqref{eq:state}, \eqref{eq:H-el}, and \eqref{eq:H-mag}, respectively, on this two-vertex state by restricting the spatial summation to $m \in \{0,1\}$.

As a final note, an important structural constraint emerges from the $\mathrm{SU}(2)$ selection rules in our basis state. The $6j$-symbols dictate that the intertwiners $\pi_m$ must satisfy the triangular inequality $|k_{m-1}-k_m| \leq \pi_m \leq k_{m-1}+k_m$. Furthermore, $\pi_m$ intertwines two identical momenta $j_m$ from the local loop, i.e. $0\leq \pi_m \leq 2 j_m$, thus $\pi_m\in \N_0$. This parity requirement forces $k_{m-1} + k_m \in \N_0$, naturally dividing the state space into two disjoint super-selection sectors: one where the connecting links is strictly $k_m\in\N_0$, and one where $k_m\in\N_0+\frac{1}{2}$. Because physical observables, such as the Hamiltonian, commute with this parity parity constraint, they cannot couple states from different sectors. The quantum dynamics are therefore confined within each individual sector, with transition matrix elements between integer and half-integer configurations identically vanishing. This dynamical decoupling allows us to isolate and independently simulate a single sector without any loss of physical information. In this work, we restrict our attention to the sector where $k_m\in \N_0+1/2$. The extension of the analysis to the other sector follows identical steps, and is therefore omitted for brevity.

\subsection{The 4-Flux Operator}
For any edges $e,e'\in\gamma_M$, the action \eqref{flux_action} of $P_I(e)$ can be extended to arbitrary Wigner matrices $D^{(j)}(g_{e'})$ as 
\begin{align}
    P_I(e)D^{(j)}(g_{e'}) = \begin{cases}
        D^{(j)}(g_{e'})\,\pi^{(j)}(\tau_I), & e' = e, \\
        -\pi^{(j)}(\tau_I)\, D^{(j)}(g_{e'}), & e' = e^{-1}, \\
        D^{(j)}(g_{e'}), & \text{otherwise}.
    \end{cases}
\end{align}
Using this expression, it can be shown that the global 4-flux operator $\O_{\rho,M}\in\L(\H_M)$ constructed from \eqref{4-flux_loc} admits matrix elements of the form
\begin{align}
    \label{4-flux_mat}
    &{}_M\!\bra{j',\pi',k'}\O_{\rho,M}\ket{j,\pi,k}_M = \left(\prod_{i\in\Z_M}\delta_{j_i j_i'}\delta_{k_i k_i'}\right) \nonumber \\
    &\times\sum_{m\in\Z_M}\left(\prod_{p\neq m}\delta_{\pi_p \pi_p'}\right) W_{j_m}^2 W_{k_{m-1}} W_{k_m} \sqrt{d_{\pi_m} d_{\pi_m'}} \nonumber \\
    & \qquad \times \left\{\begin{array}{ccc}
        j_m & 1 & j_m \\
        j_m & 1 & j_m \\
        \pi_m & \rho & \pi_m'
    \end{array}\right\}\left\{\begin{array}{ccc}
        k_{m-1} & 1 & k_{m-1} \\
        k_m & 1 & k_m \\
        \pi_m' & \rho & \pi_m
    \end{array}\right\},
\end{align}
where $W_j \coloneqq \sqrt{j(j+1)(2j+1)}$, and the $3\times 3$ arrays denote Wigner $9j$-symbols \cite{Brink:2023}. Because the triangular conditions must be satisfied across all rows and columns of $9j$-symbols, these matrix elements necessarily vanish unless $\rho\in\{0,1,2\}$. Note that this operator is diagonal in the $j$- and $k$-labels and acts non-trivially only on the intertwiner degrees of freedom.

\section{The Hamiltonian Renormalization Framework \label{app:RG}}
This appendix provides additional details on the Hamiltonian renormalization framework introduced in \Cref{sec:RG}. We review the inductive-limit construction of continuum Hilbert spaces and observables, and specialize this framework to the filling-kernel embedding \eqref{local_embed}. We then formulate the resulting operator-level \gls{RG} flow and demonstrate its application to the 4-flux operator \eqref{4-flux_loc}.


\subsection{Inductive Limits and Cylindrical Consistency \label{app:RG_inductive}}
To review the inductive-limit framework underlying the Hamiltonian \gls{RG} program, as outlined in \Cref{sec:RG}, we consider Hilbert spaces $\H_M$ indexed by finite resolution parameters $M\in\mathcal{I}$. We assume that the index set $\mathcal{I}$ is equipped with a partial order $\leq$, and that for any $M,M'\in\mathcal{I}$, there exists $M''\in\mathcal{I}$ such that $M,M'\leq M''$ \cite{Hamiltonian_RG_I, Bahr_2022}. The family $\{\H_M\}_{M\in\mathcal{I}}$ is said to be an \textit{inductive system} whenever there exist embedding maps \eqref{embed} satisfying two key properties. First, each such map is required to be an \textit{isometry}, 
\begin{align}
    \label{embed_isometry}
    \bigl\langle I_{M\to M'}\psi, I_{M\to M'}\phi\big\rangle_{\H_{M'}} = \langle \psi, \phi \rangle_{\H_M}, && M\leq M',
\end{align}
thereby ensuring that inner products are preserved under refinement. Second, the family of embeddings must be \textit{cylindrically consistent},
\begin{align}
    \label{embed_cyl}
    I_{M' \to M''}\,I_{M\to M'} = I_{M \to M''}, && M \leq M' \leq M'',
\end{align}
so that successive refinements $M \to M' \to M''$ agree with direct refinement $M \to M''$.

If $\{\H_M\}_{M\in\mathcal{I}}$ is an inductive system, then it admits an \textit{inductive limit} -- that is, there is a continuum Hilbert space $\H_\infty$ and associated embeddings $I_M:\H_M \to \H_\infty$ satisfying \cite{Bahr_2022}
\begin{align}
    I_{M'}\,I_{M\to M'} = I_M, && M \leq M'.
\end{align}
To construct $\H_\infty$, we first define an equivalence relation $\sim$ on $\{\H_M\}_{M\in\mathcal{I}}$ such that for any $\psi_M\in\H_M, \psi_{M'}\in\H_{M'}$, we have 
\begin{align}
    \psi_M \sim \psi_{M'} \iff I_{M\to M''}\psi_M = I_{M' \to M''}\psi_{M'},
\end{align}
for all $M''\in\mathcal{I}$ such that $M,M'\leq M''$ \cite{Hamiltonian_RG_I}. In other words, the equivalence class $[\psi_M]$ contains all embeddings of $\psi_M$ into higher-resolution Hilbert spaces, as well as all states in lower-resolution Hilbert spaces whose embedding into $\H_M$ yields $\psi_M$. The collection of all such equivalence classes determines the pre-Hilbert space,
\begin{align}
    \label{pre-Hilbert}
    \mathcal{D}_\infty \coloneqq \coprod_{M\in\mathcal{I}}\H_M \bigg/ \sim ,
\end{align}
which can be equipped with an inner product of the form
\begin{align}
    \bigl\langle [\psi_M], [\psi_{M'}]\bigr\rangle \coloneqq \bigl\langle I_{M\to M''}\psi_M, I_{M'\to M''}\psi_{M'}\bigr\rangle_{\H_{M''}},
\end{align}
where $M,M'\leq M''$ \cite{Hamiltonian_RG_I,Bahr_2022}. The inductive-limit Hilbert space is then obtained as the completion of \eqref{pre-Hilbert},
\begin{align}
    \label{H_inf}
    \H_\infty \coloneqq \overline{\mathcal{D}}_\infty,
\end{align}
and the continuum embedding maps are defined via
\begin{align}
    I_M:\H_M \to \H_\infty:\psi_M \mapsto [\psi_M].
\end{align}

After the continuum Hilbert space $\H_\infty$ has been obtained, one is generally interested in constructing continuum observables from families of finite-resolution operators\footnote{Although we use the term ``operators'' in this section, the discussion applies more generally to \textit{bilinear forms} defined on dense domains $\mathcal{D}_M\subset\H_M$ \cite{Bahr_2022}.}. Let $\{O_M\}_{M\in\mathcal{I}}$ be such a family, where each $O_M$ is defined on a dense domain $\mathcal{D}_M\subset \H_M$. This family determines a continuum observable $O_\infty$ on $\H_\infty$ provided its action is compatible with refinement across resolutions. In particular, we require
\begin{align}
    \label{op_cyl}
    I_{M\to M'}\,\mathcal{D}_M \subset \mathcal{D}_{M'}, && I_{M\to M'}\,O_M = O_{M'}\,I_{M\to M'},
\end{align}
where the latter condition is the operator-level realization of (\textit{strong}) \textit{cylindrical consistency} (cf. \eqref{cyl_cons_op}) \cite{Bahr_2022}. Whenever these conditions are met, we may unambiguously define the action of $O_\infty$ on $\mathcal{D}_\infty \subset \H_\infty$ via 
\begin{align}
    \label{O_inf}
    O_\infty[\psi_M] \coloneqq I_M\left(O_M\psi_M\right),
\end{align}
which, in turn, determines a bilinear form on $\H_\infty$ \cite{Hamiltonian_RG_I,Bahr_2022}.

The inductive-limit Hilbert space $\H_\infty$ and continuum observable $O_\infty$ both depend strongly on the choice of embeddings $\{I_{M\to M'}\}_{M \leq M'}$. Nevertheless, different embedding prescriptions yield inductive limits which are equivalent up to unitary transformations \cite{Hamiltonian_RG_I, Bahr_2022}. We therefore fix the embeddings according to practical considerations -- namely, feasibility of fixed-point computations -- and focus primarily on the resulting operator-level renormalization. In the following subsection, we provide further details on the specific choice of embedding employed throughout this article.

\subsection{Filling-Kernel Inductive Limit \label{app:RG_fill}}
For the remainder of this appendix, we restrict attention to the class of graphs $\{\gamma_M\}_{M\in\mathcal{I}}$, for which resolutions are indexed via $\mathcal{I} \coloneqq \{2^\ell\}_{\ell\in\N_0}$.

The aim of this subsection is to illustrate the construction of the inductive-limit Hilbert space in the case of the filling kernel \eqref{local_embed}. To this end, we construct local measures on the Hilbert spaces associated with the symmetry-reduced graph $\gamma_M$ such that $I^\text{fill}_{M\to 2M}$ satisfies the isometry condition \eqref{embed_isometry}. In particular, we work with the family of gauge-invariant Hilbert spaces $\{\H_{G,M}\}_{M\in\mathcal{I}}$, and require each embedding to satisfy 
\begin{align}
    \label{embed_isometry_gauge-inv}
    I^{\text{fill}\dagger}_{M\to 2M}\,I^\text{fill}_{M\to 2M} = \id_{\H_{G,M}},
\end{align}
with respect to the basis states \eqref{eq:state}.

Upon quantization of the symmetry-restricted phase space \eqref{M_bar}, each factor $\overline{\M}_{m,i}$ gives rise to a local $L_2$-Hilbert space of the form 
\begin{align}
    \label{H_m,i}
    \H_{m,i} \cong L_2\left(\mathrm{SU}(2),d\mu_i\right),
\end{align}
with $m\in\Z_M$ and $i\in\{x,y\}$. Inner products between functions $f_{m,i},f_{m,i}'$ on $\H_{m,i}$ are then defined with respect to the measure $\mu_i$ via
\begin{align}
    \bigl\langle f_{m,i}, f_{m,i}'\bigr\rangle_{\mu_i} &\coloneqq \mu_i\left(\overline{f}_{m,i}\,f_{m,i}'\right) \nonumber \\
    &= \int_{\mathrm{SU}(2)}\overline{f_{m,i}(g)}\,f_{m,i}'(g)\,d\mu_i(g).
\end{align}
Moreover, the measure on the kinematical Hilbert space $\H_M \cong \bigotimes_{m\in\Z_M}\left(\H_{m,x}\otimes\H_{m,y}\right)$ is given by \cite{Thiemann_MCQGR}
\begin{align}
    \label{mu_M}
    \mu_M \coloneqq \prod_{m\in\Z_M}\left(\mu_x\otimes\mu_y\right),
\end{align}
which determines inner products on $\H_M$, and hence on $\H_{G,M}$. To ensure that \eqref{embed_isometry_gauge-inv} holds, we must therefore identify $\mu_x,\mu_y$ such that 
\begin{align}
    \label{isometry_IP}
    {}^{\;\,\text{fill}}_{2M}\!\braket{j',\pi', k'|j,\pi,k}_{2M}^{\text{fill}} = {}_{M}\!\braket{j',\pi',k'|j,\pi,k}_M,
\end{align}
where 
\begin{align}
    \ket{j,\pi,k}_{2M}^{\text{fill}} \coloneqq I^\text{fill}_{M\to 2M}\ket{j,\pi,k}_M.
\end{align}

Using the explicit embedding \eqref{local_embed}, we have 
\begin{align}
    \ket{j,\pi,k}_{2M}^\text{fill} &= \ket{\left\{j_m,\pi_m,k_m,j_m,0,k_m\right\}_{m\in\Z_M}}  \\
    & = \sum_r\Bigg[\prod_{m\in\Z_M} d_{j_m}\sqrt{d_{\pi_m}}\, d_{k_m} \nonumber \\
    &\times I^\text{fill}_{M\to 2M}\left(\psi_{\pi_m}^{(k_{m-1} j_m k_m)}(h_m, g_m)\right)_{r_{m-1} r_m}\Bigg],\nonumber
\end{align}
and so \eqref{isometry_IP} reduces to a corresponding condition on the single-vertex wavefunctions \eqref{psi_vertex},
\begin{align}
    \label{psi_isometry_IP}
    &\Bigl\langle I^\text{fill}_{M\to 2M}\psi_{\pi'}^{(k_-' j' k_+')}, I^\text{fill}_{M\to 2M}\psi_{\pi}^{(k_- j k_+)}\Bigr\rangle_{(\mu_x\otimes\mu_y)^{\otimes 2}} \nonumber \\
    &\qquad= \Biggl\langle \frac{\psi_{\pi'}^{(k_-' j' k_+')}}{\sqrt{d_{j'}d_{k_+'}}}, \frac{\psi_{\pi}^{(k_- j k_+)}}{\sqrt{d_{j} d_{k_+}}}\Biggr\rangle_{\mu_x\otimes \mu_y}.
\end{align}
To this end, it can be shown that the induced local action of $I^\text{fill}_{M\to 2M}$ reads 
\begin{align}
    I^\text{fill}_{M\to 2M}\psi_{\pi}^{(k_- j k_+)}(h, g) &= \frac{\psi_{\pi}^{(k_- j k_+)}(h, g g')}{\sqrt{d_{j} d_{k_+}}}\,\mathrm{Tr}\left[D^{(j)}(h')\right],
\end{align}
where we have made use of the identity \cite{Bambi:2023jiz}
\begin{align}
    \left(\iota_0^{(jjkk)}\right)_{n_1 n_2 n_3 n_4} = \frac{(-1)^{2(j + k)}}{\sqrt{d_j d_k}}\left(\begin{array}{c}
         j  \\
         n_1\;n_2 
    \end{array}\right)\left(\begin{array}{c}
         k  \\
         n_3\;n_4 
    \end{array}\right).
\end{align}
Because the Haar measure is invariant under right-multiplication $g \mapsto g g'$ \cite{Thiemann_MCQGR}, taking $\mu_x \coloneqq \mu_H$ yields 
\begin{align}
    \label{psi_fill_IP}
    &\Bigl\langle I^\text{fill}_{M\to 2M}\psi_{\pi'}^{(k_-' j' k_+')}, I^\text{fill}_{M\to 2M}\psi_{\pi}^{(k_- j k_+)}\Bigr\rangle_{(\mu_x\otimes\mu_y)^{\otimes 2}}  \\
    &= \Biggl\langle \frac{\psi_{\pi'}^{(k_-' j' k_+')}}{\sqrt{d_{j'}d_{k_+'}}}, \frac{\psi_{\pi}^{(k_- j k_+)}}{\sqrt{d_{j} d_{k_+}}}\Biggr\rangle_{\mu_H\otimes \mu_y}\Bigl\langle\mathrm{Tr}D^{(j')},\mathrm{Tr}D^{(j)}\Bigr\rangle_{\mu_y},\nonumber
\end{align}
and it remains to determine the measure $\mu_y$ associated with the compactified edges.

Motivated by the Haar orthogonality relation \cite{Bambi:2023jiz},
\begin{align}
    \label{Haar_orthog}
    \mu_H\left(\overline{D^{(j')}_{p' q'}} D^{(j)}_{p q}\right) = \frac{\delta_{j j'}}{\sqrt{d_j d_{j'}}}\,\delta_{pp'}\delta_{qq'},
\end{align}
let us suppose that the measure $\mu_y$ satisfies
\begin{align}
    \label{mu_y_orthog}
    \mu_y\left(\overline{D^{(j')}_{p'q'}} D^{(j)}_{p q}\right) = \frac{\delta_{j j'}}{\left(d_j d_{j'}\right)^w}\, \delta_{p p'}\delta_{q q'},
\end{align}
for some exponent $w\in\mathbb{Q}$. Then \eqref{psi_fill_IP} can be evaluated explicitly, yielding
\begin{align}
    &\Bigl\langle I^\text{fill}_{M\to 2M}\psi_{\pi'}^{(k_-' j' k_+')}, I^\text{fill}_{M\to 2M}\psi_{\pi}^{(k_- j k_+)}\Bigr\rangle_{(\mu_x\otimes\mu_y)^{\otimes 2}} \nonumber \\
    &= \left(d_j d_{j'}\right)^{1/2 - w}\Biggl\langle \frac{\psi_{\pi'}^{(k_-' j' k_+')}}{\sqrt{d_{j'}d_{k_+'}}}, \frac{\psi_{\pi}^{(k_- j k_+)}}{\sqrt{d_{j} d_{k_+}}}\Biggr\rangle_{\mu_H\otimes \mu_y},
\end{align}
and we see that this reduces to \eqref{psi_isometry_IP} precisely when $w \coloneqq 1/2$. In that case, \eqref{mu_y_orthog} coincides with \eqref{Haar_orthog}, thereby allowing us to take $\mu_y \coloneqq \mu_H$.

Putting everything together, we see that \eqref{embed_isometry_gauge-inv} holds when each local Hilbert space \eqref{H_m,i} is of the form
\begin{align}
    \H_{m,i}\cong L_2\left(\mathrm{SU}(2), d\mu_H\right).
\end{align}
Thus, the filling-kernel embedding uniquely selects the standard Haar measure on both the compactified and non-compactified edges. Moreover, $\{I^\text{fill}_{M\to 2M}\}_{M\in\mathcal{I}}$ satisfies \eqref{embed_cyl} by construction, so that the choice $\mu_x = \mu_y = \mu_H$ guarantees $\{\H_{G,M}\}_{M\in\mathcal{I}}$ defines an inductive system. The continuum measure $\mu_\infty$ can then be reconstructed from the family of finite-resolution measures $\{\mu_M\}_{M\in\mathcal{I}}$ by means of the standard procedure \cite{AshtekarLewandowski1995,Hamiltonian_RG_I}:
let $F_\infty$ denote the inductive limit of a cylindrically consistent family of finite-resolution functions $\{F_M\}_{M\in\mathcal{I}}$, then \cite{Thiemann_MCQGR}
\begin{align}
    \mu_\infty(F_\infty) \coloneqq \mu_M(F_M) = \left(\prod_{m\in\Z_M}\mu_H\otimes\mu_H\right)(F_M).
\end{align}

\subsection{Operator-Level Renormalization\label{app:RG_op}}
We now illustrate the renormalization of families of finite-resolution operators $\{O_M\}_{M\in\mathcal{I}}$. In particular, our focus here is the \gls{RG} flow generated by a coarse-graining map,
\begin{align}
    \label{cg_map}
    \A_{2M \to M}\left(O_{2M}\right) \coloneqq I^\dagger_{M\to 2M}\,O_{2M}\,I_{M\to 2M},
\end{align}
and the corresponding fixed-point structure of the family of observables. To this end, we begin by introducing the notion of translation-invariant operators employed throughout this work, and then proceed to explicitly construct a fixed-point family associated with the 4-flux operator \eqref{4-flux_loc} and filling kernel \eqref{local_embed}.

\subsubsection{Translation-Invariant Operators\label{app:RG_transl_inv}}
As discussed in \Cref{sec:alg}, we consider translation-invariant operators on $\H_M$ of the form 
\begin{align}
    \label{O_M_sum}
    O_M = \sum_{m\in\Z_M} O_M(m),
\end{align}
where $O_M(m)\in\L(\H_M)$ denotes the extension of $O_m\in\L(\H^\text{loc}_m)$ to the full Hilbert space, and is given by the summand in \eqref{global_op}. 
Translation invariance means that all contributions $O_M(m)$ share the same internal structure, differing only by the subset of vertices on which they act non-trivially. Equivalently, 
\begin{align}
    O_M(m + \ell) = T_M^\ell\,O_M(m)\,T_M^{-\ell}, && \forall m,\ell\in\Z_M,
\end{align}
where
\begin{align}
    T_M^\ell:\H_M \to \H_M:\bigotimes_{m\in\Z_M}\ket{\psi_m}\mapsto \bigotimes_{m\in\Z_M}\ket{\psi_{m-\ell}}
\end{align}
is the lattice translation operator. Consequently, the global operator \eqref{O_M_sum} is completely determined by any single representative $O_M(m)$, or equivalently by the underlying local operator $O_m$.

Although the \gls{RG} flow \eqref{op_flow} can be formulated directly in terms of the local operator $O_m$, it is more convenient in the present context to work with its extension $O_M(m)$ to the full Hilbert space.
By linearity of the coarse-graining map \eqref{cg_map}, the resulting flow preserves the decomposition \eqref{O_M_sum}:
\begin{align}
    \label{O_M_flow}
    O_M^{(n)} = \sum_{m\in\Z_M}O_M^{(n)}(m).
\end{align}
The problem therefore reduces to the determination of the \gls{RG} flow $\{O_M^{(n)}(m)\}_{n\in\N_0}$.

To define the flow of $O_M(m)$, we must specify how such operators are identified under refinement $M\to 2M$. 
In the case of the filling kernel \eqref{local_embed}, each vertex $m\in\Z_M$ is replaced by a pair of neighbouring vertices $(2m,2m+1)\in\Z_{2M}^2$ on the refined lattice. Under such refinements, $O_M(m)$ is naturally identified with the corresponding sum of contributions to the resolution-$2M$ operator:
\begin{align}\label{op_refine}
    O_M(m) \xmapsto{\;M\to 2M\;} O_{2M}(2m) + O_{2M}(2m+1).
\end{align}
The resulting \gls{RG} flow is therefore defined by
\begin{align}
    \label{O_M(m)_flow}
    O_M^{(n+1)}(m) = \A_{2M \to M}\left[O_{2M}^{(n)}(2m) + O_{2M}^{(n)}(2m+1)\right],
\end{align}
from which the flow of the global operator is reconstructed using \eqref{O_M_flow}.

\subsubsection{Analytical Renormalization of the 4-Flux Operator\label{app:RG_4-flux}}
We now demonstrate the \gls{RG} procedure discussed above in the case of the 4-flux operator $\O_{\rho,M}$ and the coarse-graining map determined by the filling kernel \eqref{local_embed}. Adopting the notation of \eqref{O_M_sum}, the local operator \eqref{4-flux_loc} is extended to $\H_M$ via
\begin{align}
    \O_{\rho,M}(m) \coloneqq \left(\bigotimes_{r < m-1}\id_r\right)\otimes \O_{\rho,m}\otimes \left(\bigotimes_{s > m}\id_s\right),
\end{align}
corresponding to $\delta_0 = 1, \delta_1 = 0$ in \eqref{global_op}. Moreover, we study the \gls{RG} flow $\{\O_{\rho,M}^{(n)}\}_{n\in\N_0}$ arising from the coarse-graining map
\begin{align}
    \label{fill_CG}
    \A^\text{fill}_{2M \to M}&:\L(\H_{G,2M}) \to \L(\H_{G,M}) \nonumber \\
     &:O_{2M}\mapsto I^{\text{fill}\dagger}_{M\to 2M}\,O_{2M}\,I^\text{fill}_{M\to 2M}.
\end{align}
Because the 4-flux operator $\O_{\rho,M}$ is gauge invariant, it may be viewed as an operator on $\H_{G,M}$, so that $\A^\text{fill}_{2M \to M}$ is well-defined on the corresponding family of observables $\{\O_{\rho,M}\}_{M\in\mathcal{I}}$.

The matrix elements of $\O_{\rho,M}(m)$ in the $\ket{j,\pi,k}_M$-basis can be read off directly from \eqref{4-flux_mat} by simply taking the term associated with $m\in\Z_M$. It can then be shown that
\begin{equation}\label{O_rho_flow_terms}
    \begin{gathered}
        \A^\text{fill}_{2M \to M}\O_{\rho, 2M}^{(n)}(2m) = \O_{\rho,M}^{(n)}(m), \\
        \A^\text{fill}_{2M \to M}\O_{\rho, 2M}^{(n)}(2m+1) = 2^n\,\delta_{\rho,0}\,\Lambda^\text{fill}_M(m),
    \end{gathered}
\end{equation}
where $\Lambda^\text{fill}_M(m)$ is characterized by 
\begin{align}
    &{}_{M}\!\bra{j',\pi',k'}\Lambda^\text{fill}_M(m)\ket{j,\pi,k}_M = \frac{1}{3}\left(\prod_{i\in\Z_M}\delta_{j_i j_i'}\delta_{\pi_i \pi_i'}\delta_{k_i k_i'}\right) \nonumber \\ 
    &\times j_m\big(j_m + 1\big)k_m\big(k_m + 1\big)\big\{j_m\;1\;j_m\big\}\big\{k_m\;1\;k_m\big\},
\end{align}
with $\{j_1\;j_2\;j_3\}$ simply enforcing the triangular conditions between $j_1,j_2,j_3$. 

Combining \eqref{O_rho_flow_terms} with \eqref{O_M(m)_flow}, we obtain
\begin{align} 
    \label{4-flux_filling}
    \O_{\rho,M}^{(n)}(m) = \O_{\rho,M}(m) + \big(2^n - 1\big)\,\delta_{\rho,0}\,\Lambda^\text{fill}_M(m),
\end{align}
so that all dependence on the \gls{RG} step $n$ is contained in the prefactor of the second term. This expression shows that the deviation of $\O_{\rho,M}(m)$ from fixed-point behaviour under $\A^\text{fill}_{2M\to M}$ is governed entirely by $\Lambda_M^\text{fill}(m)$.

Equation \eqref{4-flux_filling} also implies that the limit $\lim_{n\to\infty}\O_{\rho,M}^{(n)}(m)$ does not exist for $\rho = 0$. Thus, the \gls{RG} flow generated by the filling kernel does not converge to a fixed point in general. Nevertheless, a fixed point associated with $\O_{\rho,M}(m)$ can still be constructed by observing that $\Lambda_M^\text{fill}(m)$ transforms according to 
\begin{align}
    \label{lambda_step}
    \A^\text{fill}_{2M \to M}\left[\Lambda^\text{fill}_{2M}(2m) + \Lambda^\text{fill}_{2M}(2m+1)\right] = 2\,\Lambda^\text{fill}_M(m).
\end{align}
It follows immediately that 
\begin{align} 
    \label{4-flux_FP}
    \O_{\rho, M}^*(m) \coloneqq \O_{\rho, M}(m) - \delta_{\rho, 0}\,\Lambda^\text{fill}_M(m)
\end{align}
defines a fixed point of the coarse-graining map $\A^\text{fill}_{2M \to M}$. Note that for $\rho \neq 0$ the correction term vanishes, so that $\O_{\rho,M}^*(m)$ coincides precisely with the bare operator. The associated global fixed point can then be reconstructed via 
\begin{align}
    \label{4-flux_FP_global}
    \O_{\rho,M}^* \coloneqq \sum_{m\in\Z_M}\O_{\rho,M}^*(m).
\end{align}

\section{VAPOR: Background and Construction \label{app:alg}}
This appendix develops the Pauli-level renormalization framework underlying VAPOR (see \Cref{sec:alg}). Our goal is to reformulate the operator-level \gls{RG} analysis of locally-defined, translation-invariant operators in terms of induced \gls{RG} flows on individual Pauli strings. This reduces the fixed-point problem to a collection of finite-dimensional eigenvalue problems on \gls{RG}-invariant Krylov-like subsectors.


\subsection{Pauli Decompositions of Translation-Invariant Operators\label{app:alg_pauli-decomp}}

Suppose that the label data $\{(j_m,\pi_m,k_m)\}_{m\in\Z_M}$ associated with a state at resolution $M$ is represented using a truncation with $q\in\N$ qubits per label, for a total of $Q\coloneqq 3Mq$ qubits. Then each label type $j,\pi,k$ admits $2^q$ distinct values\footnote{In general, one may choose different qubit counts $q_I\in\N$ for each label type $I\in\{j,\pi,k\}$. For simplicity, however, we restrict to the special case $q_I = q,\,\forall I$ throughout this appendix.}, allowing one to naturally define, e.g.,
\begin{align}
    j_\text{max} = \frac{1}{2}\left(2^q - 1\right).
\end{align}
The resulting $Q$-qubit Hilbert space, $\H_Q \coloneqq (\C^2)^{\otimes Q}$, contains a truncation of the gauge-invariant Hilbert space $\H_{G,M}$ as a  subspace, which can be uniquely identified upon associating representation data with computational-basis (i.e., qubit) states. Although $\H_{G,M}$ does not generally admit a tensor-product factorization \cite{Bambi:2023jiz}, the associated qubit Hilbert space factorizes as 
\begin{align}
    \label{H_Q_decomp}
    \H_Q \cong \bigotimes_{m\in\Z_M}\H_{3q}, && \H_{3q} \coloneqq \left(\C^2\right)^{\otimes 3q},
\end{align}
where each factor $\H_{3q}$ is associated with the three labels $(j_m,\pi_m,k_m)$ at a given vertex $m\in\Z_M$. Accordingly, the Pauli basis on $\H_Q$ can be expressed as
\begin{align}
    \P_Q \cong \bigotimes_{m\in\Z_M}\P_{3q}, && \P_{3q} \coloneqq \{I,X,Y,Z\}^{\otimes 3q},
\end{align}
where $|\P_Q| = 4^Q$, and $\P_{3q}$ denotes the set of Pauli strings acting on $\H_{3q}$ \cite{NielsenChuang}.

Here, we assume that the representation data $(j,\pi,k)$ associated with a single vertex are encoded via 
\begin{align}
    \label{local_encode}
    (j,\pi,k)\longmapsto \ket{j}_q\otimes\ket{\pi}_q\otimes\ket{k}_q,
\end{align}
where $\ket{j}_q,\ket{\pi}_q,\ket{k}_q\in(\C^2)^{\otimes q}$ denote computational-basis states associated with the corresponding label values. With this choice, the single-vertex Hilbert space $\H_{3q}$ further decomposes into three independent $q$-qubit registers -- one for each label type:
\begin{align}
    \label{H_3q_label_decomp}
    \H_{3q} \cong \bigotimes_{I\in\{j,\pi,k\}}\H_q^{(I)}, && \H_q^{(I)} \cong \left(\C^2\right)^{\otimes q}.
\end{align}
It follows that any Pauli string $P\in\P_{3q}$ admits a unique decomposition of the form 
\begin{align}
    P = P_j\otimes P_\pi \otimes P_k\in\P_q^{\otimes 3},
\end{align}
where $P_I\in\P_q$ acts only on the qubits in $\H_q^{(I)}$, $\forall I\in\{j,\pi,k\}$.
Moreover, applying \eqref{local_encode} to each vertex $m\in\Z_M$ yields the global encoding
\begin{align}
    \label{global_encode}
    \left\{(j_m,\pi_m,k_m)\right\}_{m\in\Z_M}\longmapsto \bigotimes_{m\in\Z_M}\ket{j_m,\pi_m,k_m}_{3q},
\end{align}
where $\ket{j_m,\pi_m,k_m}_{3q} \coloneqq \ket{j_m}_q\otimes\ket{\pi_m}_q\otimes\ket{k_m}_q\in\H_{3q}$. Restricting \eqref{global_encode} to the label values associated with the basis states \eqref{eq:state} therefore yields a qubit-level truncation of $\H_{G,M}$ within $\H_Q$.

We characterize the Pauli structure of translation-invariant operators of the form \eqref{O_M_sum}. A representative term $O_M(m)\in\L(\H_M)$ acts non-trivially only on the local Hilbert space $\H^\text{loc}_m = \H_{m-\delta_0}\otimes \cdots \otimes \H_{m+\delta_1}$ associated with $\delta\coloneqq \delta_0 + \delta_1 + 1$ vertices. The corresponding qubit-encoded Hilbert space and local Pauli basis are given by
\begin{align}
    \H_{3\delta q} \cong \bigotimes_{1\leq i \leq \delta}\H_{3q}, && \P_{3\delta q} \cong \bigotimes_{1\leq i \leq \delta}\P_{3q}.
\end{align}

Assuming $\delta \leq M$, one may always choose $m\in\Z_M$ such that $m - \delta_0 \geq 0$ and $m + \delta_1 \leq M - 1$\footnote{
    If the local operator's vertex support $\{m-\delta_0,\ldots,m+\delta_1\}$ wraps around the periodic lattice, the same construction applies after re-interpreting all vertex labels modulo $M$.
}, in which case $O_M(m)$ admits a decomposition in the global Pauli basis $\P_Q$ of the form 
\begin{align}
    \label{O_M(m)_decomp}
    O_M(m) &= \sum_\alpha c_\alpha(M)\,I^{\otimes 3(m - \delta_0)q}\otimes P_\alpha \nonumber \\
    &\qquad\qquad \otimes I^{\otimes 3\left[M - (m+\delta_1 + 1)\right]q},
\end{align}
with each $P_\alpha\in\P_{3\delta q}$ acting on the local Hilbert space $\H_{3\delta q}$. The corresponding local operator $O_m\in\L(\H^\text{loc}_m)$ therefore decomposes as 
\begin{align}
    \label{O_m_decomp}
    O_m = \sum_\alpha c_\alpha(M)\,P_\alpha\in\Span\P_{3\delta q},
\end{align}
with coefficients inherited directly from \eqref{O_M(m)_decomp}.

If, on the contrary, $\delta >M$, we would be considering operators that should be thought of as being defined on a larger lattice than $M$. In fact, there exist a natural projection map $\L(\H_{2M})\to\L(\H_M)$ that takes the periodic boundary condition into account to map the long-range interactions back to the lattice $\mathbb{Z}_M$. The consequence of including these long-range operators will be a more involved action of the coarse-graining map, than what we will utilize in the following sections. Yet, the method is still constructive and computable and we will refer to this extension when building the fixed points of the Kogut-Susskind Hamiltonian \eqref{eq:KS-Ham} in forthcoming research.

\subsection{Pauli-Level Embedding and Coarse-Graining Maps \label{app:alg_pauli-cg}}
Recall from \Cref{app:RG} that we are interested in embedding maps $I_{M\to 2M}:\H_{G,M}\to\H_{G,2M}$ associated with vertex-level refinements of the form $m\mapsto (2m,2m+1)$. Such refinements act independently at each vertex and determine corresponding refinements of the associated representation labels. In particular, if $(j,\pi,k)$ denotes the label data at a single vertex, then the encoding \eqref{local_encode} yields a local qubit-level embedding,
\begin{align}
    \label{I_3q_6q}
    I_{3q\to 6q}&:\H_{3q}\to \H_{3q}\otimes\H_{3q} \nonumber \\
    &: \ket{j,\pi,k}_{3q}\mapsto \ket{j_L,\pi_L,k_L}_{3q}\otimes\ket{j_R,\pi_R,k_R}_{3q},
\end{align}
where $\ket{j_L,\pi_L,k_L}_{3q},\ket{j_R,\pi_R,k_R}_{3q}$ correspond to the left and right vertices in the refined pair, respectively. Because the refinement acts independently at each $m\in\Z_M$, the tensor-product structure \eqref{H_Q_decomp} of $\H_Q$ induces a global embedding map,
\begin{align}
    I_{Q\to 2Q}:\H_Q \to \H_{2Q},
\end{align}
defined by
\begin{align}
    \label{I_Q_2Q}
    I_{Q \to 2Q} \coloneqq \bigotimes_{m\in\Z_M}I_{3q \to 6q}.
\end{align}

Each embedding $I_{Q\to 2Q},I_{3q \to 6q}$ defines an associated coarse-graining map by conjugation (see \eqref{cyl_cons_op}). 
Pauli decompositions naturally reframe operators as vectors in the linear span of the corresponding Pauli basis, in which case coarse-graining maps serve to identify refined vectors with their coarse counterparts.
In particular, the local embedding $I_{3q \to 6q}$ yields a map 
\begin{align}
    \A_{6q \to 3q}:\Span\P_{6q}\to\Span\P_{3q},
\end{align}
acting on two-vertex Pauli strings $P_L\otimes P_R\in\P_{6q}$ according to 
\begin{align}
    \label{A_6q_3q}
    \A_{6q \to 3q}\left(P_L\otimes P_R\right) \coloneqq I_{3q\to 6q}^\dagger\left(P_L\otimes P_R\right) I_{3q\to 6q}.
\end{align}
Due to the tensor-product structure \eqref{I_Q_2Q} of the global embedding, the induced coarse-graining map on $\Span\P_{2Q}$ can be reconstructed from \eqref{A_6q_3q} via
\begin{align}
    \label{A_2Q_Q}
    \A_{2Q \to Q} = \bigotimes_{m\in\Z_M}\A_{6q \to 3q},
\end{align}
which is the direct analog of the operator-level coarse-graining map $\A_{2M \to M}$.

Now, depending on the precise structure of the chosen embedding, it is sometimes possible to further reduce the action \eqref{A_6q_3q}. For example, in the case of the filling kernel \eqref{local_embed}, the local qubit-level embedding is given by
\begin{align}
    \label{I_fill_loc}
    I^\text{fill}_{3q \to 6q}\ket{j,\pi,k}_{3q} &= \left(\ket{j}_q\otimes\ket{\pi}_q\otimes\ket{k}_q\right) \nonumber \\
    &\qquad \otimes\left(\ket{j}_q\otimes\ket{0}_q\otimes\ket{k}_q\right),
\end{align}
where $\ket{0}_q\in(\C^2)^{\otimes q}$ denotes the all-zero computational-basis state on a single $q$-qubit register. Because this map does not mix different label sectors, the associated coarse-graining map factorizes into a set of independent, label-specific maps:
\begin{align}
    \label{A_fill_loc}
    \A^\text{fill}_{6q \to 3q} = \bigotimes_{I\in\{j,\pi,k\}}\A^{\text{fill}(I)}_{2q \to q},
\end{align}
where $\A^{\text{fill}(I)}_{2q\to q}$ acts only on the Pauli-factors associated with the label $I$ (recall \eqref{H_3q_label_decomp}).

Moreover, as no new non-trivial $q$-qubit states appear on the right-hand side of \eqref{I_fill_loc}, each map $\A^{\text{fill}(I)}_{2q\to q}$ can be further decomposed into a bitwise product across single-qubit states. Explicitly, given Pauli strings $P_{I_L},P_{I_R}\in\P_q$ associated with the $I$-label sector, we can write 
\begin{align}
    P_{I_L} = \bigotimes_{1\leq w \leq q}P_{I_L,w}, && P_{I_R} = \bigotimes_{1\leq w \leq q}P_{I_R,w},
\end{align}
and one has 
\begin{align}
    \label{A_fill_label}
    \A^{\text{fill}(I)}_{2q \to q}\left(P_{I_L}\otimes P_{I_R}\right) = \bigotimes_{1\leq w \leq q}\A^{\text{fill}(I)}_{2\to 1}\left(P_{I_L,w}\otimes P_{I_R,w}\right).
\end{align}
In this way, the multi-qubit coarse-graining map induced by the filling kernel is fully determined by a set of basic actions on $\P_1\otimes\P_1$. In \Cref{app:alg_4-flux}, we will return to this specific coarse-graining map and construct its action explicitly.

\subsection{Induced RG Map on Local Pauli Operators \label{app:alg_pauli-RG}}
It is apparent from \eqref{I_Q_2Q} and \eqref{A_2Q_Q} that the local maps $I_{3q \to 6q}$ and $\A_{6q \to 3q}$ extend naturally to states and operators defined on multi-vertex Hilbert spaces. Of particular interest to us is the coarse-graining map relevant to the local operator \eqref{O_m_decomp}, namely,
\begin{align}\label{A_delta}
    \A_{6\delta q \to 3\delta q} = \bigotimes_{1\leq i \leq \delta}\A_{6q \to 3q}.
\end{align}
Once a prescription for refinement of local Pauli operators under $\H_{3\delta q}\to \H_{6\delta q}$ has been fixed, this coarse-graining map therefore determines the local \gls{RG} flow $\{O_m^{(n)}\}_{n\in\N_0}$.

Recall from \Cref{app:RG} that under refinement $M \to 2M$, the operator $O_M(m)$ was chosen to transform according to \eqref{op_refine}. 
At the level of local Pauli strings $P_\alpha\in\P_{3\delta q}$, this becomes
\begin{align}
    \label{pauli_refine}
    P_\alpha \xmapsto{\;3\delta q \to 6\delta q\;} \;&I^{\otimes 3\delta_0 q}\otimes P_\alpha \otimes I^{\otimes 3(\delta_1 + 1)q} \nonumber \\
    & + I^{\otimes 3(\delta_0 + 1)q}\otimes P_\alpha \otimes I^{\otimes 3\delta_1 q},
\end{align}
with each string on the right-hand side belonging to $\P_{6\delta q}$. 
Applying the coarse-graining map \eqref{A_delta} to this refinement thus defines a linear \gls{RG} map on local Pauli operators,
\begin{align}
    \label{pauli_op_RG_map}
    \A:\Span\P_{3\delta q} \to \Span\P_{3\delta q}.
\end{align}

Now, the Pauli strings $\P_{3\delta q}$ form an orthonormal basis for $\Span\P_{3\delta q}$ with respect to the normalized Hilbert-Schmidt inner product \cite{NielsenChuang},
\begin{align}\label{HS_IP}
    \bigl\langle O, O'\bigr\rangle_{3\delta q} \coloneqq 2^{-3\delta q}\,\mathrm{Tr}\left(O^\dagger O'\right).
\end{align}
The induced \gls{RG} map \eqref{pauli_op_RG_map} therefore acts on the local Pauli basis as
\begin{align}
    \A\left(P_\beta\right) = \sum_\alpha A_{\alpha\beta} P_\alpha,
\end{align}
with the \textit{induced \gls{RG} matrix} $A$ defined explicitly by
\begin{align}\label{RG_mat}
    A_{\alpha\beta} \coloneqq \bigl\langle P_\alpha,\A\left(P_\beta\right)\bigr\rangle_{3\delta q}.
\end{align}

Finally, we consider the local \gls{RG} flow $\{O_m^{(n)}\}_{n\in\N_0}$ generated by \eqref{pauli_op_RG_map}, as well as the form of fixed points with respect to this map. 
For resolution-dependent operators of the form \eqref{O_m_decomp}, we assume that each Pauli coefficient admits an expansion of the form
\begin{align}\label{coeff_expansion}
    c_\alpha(M) = \sum_{\ell\in\Z}\frac{c_{\ell,\alpha}}{M^\ell},
\end{align}
for some coefficients $\{c_{\ell,\alpha}\}_{\ell\in\Z}$. 
Under refinement, the coefficients transform as $c_\alpha(M) \to c_\alpha(2M)$, so that the $n^\text{th}$ element of the local flow is given by
\begin{align}
    \label{pauli_op_RG_flow}
    \A^n\left(O_m\right) &= \sum_\alpha c_\alpha\left(2^n M\right)\A^n(P_\alpha) \nonumber \\
    &= \sum_{\ell\in\Z}\frac{1}{\left(2^n M\right)^\ell}\sum_\alpha\left(A^n c_\ell\right)_\alpha P_\alpha,
\end{align}
where $c_\ell\in\C^{\dim\P_{3\delta q}}$ denotes the vector with components $(c_\ell)_\alpha = c_{\ell,\alpha}$.
Moreover, any fixed point $O_m^* = \sum_\alpha c_\alpha^*(M) P_\alpha$ of \eqref{pauli_op_RG_map} necessarily satisfies 
\begin{align}
    O_m^* = \A\left(O_m^*\right) \iff \sum_{\ell\in\Z}\frac{1}{M^\ell}\left(c_\ell^* - \frac{1}{2^\ell} A\,c_\ell^*\right) = 0.
\end{align}
In particular, this condition is met precisely when
\begin{align}\label{fp_eigen}
    A\,c_\ell^* = 2^\ell c_\ell^*, && \forall \ell\in\Z,
\end{align}
i.e., each non-zero $c_\ell^*$ is an eigenvector of $A$ with eigenvalue $\lambda = 2^\ell$.

The preceding observation effectively reframes fixed points of the \gls{RG} map \eqref{pauli_op_RG_map} as eigenvectors of the induced \gls{RG} matrix \eqref{RG_mat}. The fixed-point problem thus reduces to the spectral decomposition of $A$, together with projections of the bare operator onto each relevant eigenspace. In practice, this characterization allows fixed points to be determined directly from the spectral properties of the \gls{RG} matrix, without construction of the full \gls{RG} flow $\{O_m^{(n)}\}_{n\in\N_0}$. This is particularly useful in cases where $\lim_{n\to\infty}O_m^{(n)}$ does not exist, and the \gls{RG} flow therefore fails to converge to a fixed point.

\subsection{Pauli Supports and Orbits \label{app:alg_supp-orb}}
Although the linear-algebraic formulation of the fixed-point problem simplifies matters, its direct implementation requires diagonalizing the $4^{3\delta q}\times 4^{3\delta q}$ matrix \eqref{RG_mat} -- a computationally strenuous task even for moderate values of $\delta$ and $q$. To further reduce computational complexity, we exploit the structure of the \gls{RG} map $\A$ to decompose the local Pauli basis into \gls{RG}-invariant subsectors relevant to the operator under consideration. As we will see, these subsectors naturally determine Krylov-type subspaces associated with individual Pauli strings, allowing the underlying eigenvalue problem to be reformulated as a collection of smaller, decoupled sub-problems \cite{Saad2011}.

A first simplification arises from the sparsity of many translation-invariant operators in the local Pauli basis. E.g., the 4-flux operator \eqref{4-flux_loc} admits a Pauli decomposition containing at most $4^{\delta q} \ll |\P_{3\delta q}|$ non-zero terms. Rather than considering all of $\P_{3\delta q}$, it suffices to study the subset of Pauli strings appearing in the decomposition, together with their images under the \gls{RG} map.

To this end, we first define the \textit{Pauli support} of a local operator $O_m\in\Span\P_{3\delta q}$ by
\begin{align}\label{pauli_support}
    \Supp O_m \coloneqq \big\{P_\alpha\,\big|\,\bigl\langle P_\alpha, O_m\bigr\rangle_{3\delta q}\neq 0\big\} \subset \P_{3\delta q}.
\end{align}
In other words, $\Supp O_m$ is simply the set of local Pauli strings appearing with non-zero coefficients in the decomposition \eqref{O_m_decomp}. Accounting for the expansion \eqref{coeff_expansion} of the Pauli coefficients, the decomposition \eqref{O_m_decomp} can therefore be expressed as
\begin{align}
    \label{O_m_supp_decomp}
    O_m = \sum_{\ell\in\Z}\frac{1}{M^\ell}\left(\sum_{P_\alpha\in\Supp O_m} c_{\ell,\alpha}\, P_\alpha\right).
\end{align}
Moreover, the definition \eqref{pauli_support} extends naturally to sets of operators via
\begin{align}
    \label{set_support}
    \Supp\S \coloneqq \bigcup_{O\in\S}\Supp O, && \S \subset \Span\P_{3\delta q}.
\end{align}
Similarly, the support of a subspace $\Span\S \subset \Span\P_{3\delta q}$ is denoted by $\Supp\Span\S$ and defined by applying \eqref{set_support} to any choice of basis.

While the support \eqref{pauli_support} characterizes the Pauli structure of $O_m$ itself, it is generally insufficient for describing the \gls{RG} images $\A^n(O_m)$. Indeed, the Pauli-level action \eqref{pauli_RG_map} implies that $\A(P_\alpha)\in\Span\P_{3\delta q}$ for a given string $P_\alpha\in\Supp O_m$, so that neither $\A(P_\alpha)$ nor $\Supp\A(P_\alpha)$ are necessarily contained in the original support. To systematically account for these additional contributions, we will now consider the \gls{RG} flow $\{\A^n(P_\alpha)\}_{n\in\N_0}$ of a single Pauli string.

Since each iterate $\A^n(P_\alpha)$ belongs to the finite-dimensional vector space $\Span\P_{3\delta q}$, the sequence $\{\A^n(P_\alpha)\}_{n\in\N_0}$ cannot remain linearly independent indefinitely. Hence, there exists a minimal integer $n_\alpha \leq \dim\Span\P_{3\delta q}$ -- called the \textit{grade} of $P_\alpha$ -- such that 
\begin{align}
    \A^{n_\alpha}(P_\alpha) = \sum_{n < n_\alpha}\theta_n \A^n(P_\alpha),
\end{align}
for some coefficients $\{\theta_n\}_{n < n_\alpha}$ \cite{Saad2011}. As a consequence, we see that the flow of $P_\alpha$ is completely contained within the (order-$n_\alpha$) \textit{Krylov subspace} \cite{Krylov1931},
\begin{align}
    \label{krylov}
    \K_{n_\alpha}(\A,P_\alpha)\coloneqq \Span\left\{\A^n(P_\alpha)\right\}_{n < n_\alpha}.
\end{align}
By construction, this subspace satisfies 
\begin{align}
    \A\left(\K_{n_\alpha}(\A,P_\alpha)\right) \subset \K_{n_\alpha}(\A, P_\alpha),
\end{align}
and can therefore be interpreted as an \gls{RG}-invariant vector space associated with $P_\alpha$ \cite{Saad2011}.

The set of all Pauli strings relevant to the \gls{RG} flow of $P_\alpha$ is given by the support of the associated Krylov subspace \eqref{krylov}. We therefore define the (\textit{forward}) \textit{Pauli orbit} of $P_\alpha$ with respect to $\A$ by
\begin{align}
    \label{pauli_orbit}
    \Orb_\A P_\alpha \coloneqq \Supp\K_{n_\alpha}(\A,P_\alpha) = \bigcup_{n < n_\alpha}\Supp \A^n(P_\alpha).
\end{align}
Every such orbit is necessarily finite, with the size being constrained by 
\begin{align}
    n_\alpha \leq |\Orb_\A P_\alpha| \leq |\P_{3\delta q}|.
\end{align}
Moreover, $\Orb_\A P_\alpha$ satisfies 
\begin{align}
    \Supp\A\left(\Orb_\A P_\alpha\right) \subset \Orb_\A P_\alpha,
\end{align}
thereby defining the smallest \gls{RG}-invariant Pauli support containing $P_\alpha$. Using standard techniques from Krylov theory (e.g., Arnoldi's method \cite{Arnoldi_1951}), an explicit orthonormal basis for $\K_{n_\alpha}(\A,P_\alpha)$ can be constructed from the corresponding orbit \eqref{pauli_orbit}. In this way, the orbit construction described here provides a natural bridge between \gls{RG} analyses of local Pauli operators and the well-established Krylov subspace methods employed extensively in the study of linear systems \cite{Saad2011}.

Having identified the \gls{RG}-invariant subsectors associated with each $P_\alpha\in\Supp O_m$, we now return to the fixed-point problem. 
To this end, we first recall from \eqref{fp_eigen} that any fixed point of $\A$ can be decomposed in terms of $\lambda=2^\ell$-eigenvectors of the induced \gls{RG} matrix \eqref{RG_mat}. In particular, a fixed point $O_m^*$ associated with $O_m$ can be expressed in the general form
\begin{align}
    \label{fp_sum}
    O_m^* = \sum_{2^\ell\in\sigma(A)}\frac{1}{M^\ell}\left(\sum_{P_\alpha\in\Supp O_m}c_{\ell,\alpha}\,O_{\ell,\alpha}\right),
\end{align}
where $\sigma(A)$ denotes the spectrum of $A$, and each operator $O_{\ell,\alpha}$ satisfies $\A(O_{\ell,\alpha}) = 2^\ell\,O_{\ell,\alpha}$. 

Since the flow of $P_\alpha\in\Supp O_m$ is confined to the \gls{RG}-invariant vector space \eqref{krylov}, we restrict our attention to fixed-point contributions $\{O_{\ell,\alpha}\}_{2^\ell\in\sigma(A)} \subset \K_{n_\alpha}(\A, P_\alpha)$.
More precisely, each $O_{\ell,\alpha}$ is required to solve the eigenvalue problem determined by restriction of the \gls{RG} map to $\K_{n_\alpha}(\A, P_\alpha)$:
\begin{align}
    \label{krylov_eig}
    \A\big|_{\K_{n_\alpha}(\A,P_\alpha)}\left(O_{\ell,\alpha}\right) = 2^\ell\,O_{\ell,\alpha}.
\end{align}
It follows immediately that 
\begin{align}
    \label{orb_fp}
    O_{\ell,\alpha}^* \coloneqq \frac{1}{M^\ell}\,O_{\ell,\alpha}
\end{align}
defines a fixed point of the restricted \gls{RG} map, hence of the full map $\A$ as well.

We refer to the operator \eqref{orb_fp} as an \textit{orbit-supported fixed point} associated with $P_\alpha$, owing to the fact that $\Supp O_{\ell,\alpha}^* \subset \Orb_\A P_\alpha$. 
Thus, rather than explicitly constructing a basis for the Krylov subspace $\K_{n_\alpha}(\A,P_\alpha)\subset \Span\Orb_\A P_\alpha$, we may work directly with the Pauli basis provided by the orbit $\Orb_\A P_\alpha$ itself.
To this end, let us denote by $\A_\alpha$ the restriction of the full \gls{RG} map to the vector space determined by $\Orb_\A P_\alpha$, i.e.,
\begin{align}
    \A_\alpha \coloneqq \A\big|_{\Span\Orb_\A P_\alpha}.
\end{align}
The action of $\A_\alpha$ on $P_j\in\Orb_\A P_\alpha$ precisely mirrors that of $\A$ on the local Pauli basis -- that is, 
\begin{align}
    \label{orb_RG_map}
    \A_\alpha(P_j) = \sum_{P_i\in\Orb_\A P_\alpha}\left(A_\alpha\right)_{ij} P_i,
\end{align}
with the reduced \gls{RG} matrix $A_\alpha$ given by
\begin{align}
    \label{orb_RG_mat}
    \left(A_\alpha\right)_{ij} \coloneqq \bigl\langle P_i, \A(P_j)\bigr\rangle_{3\delta q}.
\end{align}
In this case, the fixed-point contribution associated with $P_\alpha$ can be canonically defined via
\begin{align}
    \label{orb_fp_contribution}
    O_{\ell,\alpha} \coloneqq \Proj_{\ker(\A_\alpha - 2^\ell \id)}(P_\alpha),
\end{align}
where $\ker(\A_\alpha - 2^\ell\id)$ is the $\lambda = 2^\ell$-eigenspace determined by \eqref{orb_RG_map}, and the projection of $P_\alpha$ onto this subspace is computed using the Hilbert-Schmidt inner product \eqref{HS_IP}. 
Finally, as a non-trivial operator of the form \eqref{orb_fp_contribution} exists only if $2^\ell$ is an eigenvalue of the reduced matrix \eqref{orb_RG_mat}, the fixed point \eqref{fp_sum} is reconstructed as
\begin{align}
    \label{fp_orb_sum}
    O_m^* = \sum_{P_\alpha\in\Supp O_m}\left(\sum_{2^\ell\in\sigma(A_\alpha)}\frac{c_{\ell,\alpha}}{M^\ell}\,O_{\ell,\alpha}\right),
\end{align}
where the inner sum now runs over the spectrum $\sigma(A_\alpha)$ associated with a specific orbit. 

It is important to distinguish the projection-based construction from direct solutions of the homogeneous fixed-point equation. As noted above, the definition \eqref{orb_fp_contribution} produces a canonical eigenvector $O_{\ell,\alpha}\in\ker(\A_\alpha - 2^\ell\id)$, determined uniquely by means of projection. By contrast, solving the eigenvalue problem
\begin{align}
    \label{orb_eig}
    \A_\alpha\left(O_{\ell,\alpha}\right) = 2^\ell\,O_{\ell,\alpha}
\end{align}
directly within a given orbit identifies $O_{\ell,\alpha}$ only up to overall normalization. Consequently, \eqref{orb_fp_contribution} simply selects a particular representative \eqref{fp_orb_sum} from an entire family of fixed-point solutions. In practice, it is often more convenient to define the fixed-point contribution $O_{\ell,\alpha}\in\Span\Orb_\A P_\alpha$ directly as
\begin{align}
    O_{\ell,\alpha} = \sum_{P_i\in\Orb_\A P_\alpha}\theta_i\,P_i,
\end{align}
with coefficients subject to a predetermined constraint singling out a unique solution of \eqref{orb_eig}. This is the perspective we adopt in the following subsection, where the Pauli-level fixed-point construction is recast as a language-agnostic variational optimization algorithm. 

\subsection{Quantum Fixed-Point Algorithm \label{app:alg_algorithm}}
For sufficiently simple operators, the eigenspace projection procedure above can be implemented directly on a classical computer. In general, however, the orbit dimensions may approach $|\P_{3\delta q}|$, leaving the orbit-restricted \gls{RG} matrices $A_\alpha$ prohibitively large. Evaluating the projection \eqref{orb_fp_contribution} then still requires expensive eigenvalue problems, nullifying the complexity reduction that motivated the orbit-based approach.

To address this, we introduce a variational implementation of the fixed-point construction for quantum devices. Rather than explicitly diagonalizing the \gls{RG} matrices, our method reformulates the fixed-point problem as a constrained optimization over orbit-supported ans\"atze. This solves the relevant eigenvalue problems efficiently within \gls{RG}-invariant Krylov sectors, while avoiding the deep circuits and large-scale spectral decompositions typically required by quantum matrix-diagonalization methods \cite{gilyen2018qsvt,low2017optimal}. The resulting implementation is thus compatible with shallow-circuit quantum architectures and should be more robust against noise on near-term devices (see \Cref{app:VQE}) \cite{Peruzzo2014,McClean2016,Cerezo:2020jpv}.

As before, let $O_m^{(0)}$ denote a resolution-dependent local operator acting on $\H^\text{loc}_m = \H_{m-\delta_0}\otimes \cdots \otimes \H_{m+\delta_1}$. We assume that this operator admits a Pauli decomposition of the form
\begin{align}
    \label{bare_op_decomp}
    O_m^{(0)} = \sum_{\ell\in L}\frac{1}{M^\ell}\left(\sum_\alpha c_{\ell,\alpha}^{(0)}\,P_\alpha\right)\in\Span\P_{3\delta q},
\end{align}
with $c_{\ell,\alpha}^{(0)}\neq 0$ only for finitely many expansion orders $\ell\in L\subset \Z$. Moreover, rather than constructing the full orbits \eqref{pauli_orbit} for each $P_\alpha\in\Supp O_m^{(0)}$ \textit{a priori}, we instead introduce a finite truncation parameter $n_0 \leq |\P_{3\delta q}|$ and work within the order-$n_0$ orbits,
\begin{align}
    \label{n_0-orbit}
    \Orb_\A^{n_0}P_\alpha \coloneqq \Supp \K_{n_0}(\A,P_\alpha).
\end{align}
As $\dim\K_{n_0}(\A,P_\alpha) = \min\{n_0,n_\alpha\}$, where $n_\alpha$ is the grade of $P_\alpha$, the truncation parameter $n_0$ directly controls the dimension of the variational search space associated with $P_\alpha$ \cite{Saad2011}.

If no fixed point is identified within the order-$n_0$ Krylov sector, we enlarge the search space using the nested structure,
\begin{align}
    \K_{n_0}(\A, P_\alpha) \subset \K_{n_0 + 1}(\A, P_\alpha),
\end{align}
together with the corresponding orbit relation
\begin{align}
    \label{(n_0+1)-orbit}
    \Orb_\A^{n_0 + 1}P_\alpha = \left(\Orb_\A^{n_0}P_\alpha\right)\cup \Supp \A^{n_0}(P_\alpha).
\end{align}
This procedure may be repeated until the order reaches the grade $n_\alpha$, at which point further iterations no longer enlarge the underlying Krylov subspace \cite{Saad2011}, hence
\begin{align}
    \label{orbit_grade}
    \Orb_\A^{n_\alpha + 1}P_\alpha = \Orb_\A^{n_\alpha}P_\alpha.
\end{align}

\begin{alg}[VAPOR]\label{alg:vapor} The following procedure accommodates the efficient construction of fixed points associated with local operators of the form \eqref{bare_op_decomp}.
    \begin{enumerate}
        \item Construct a set of reference states $\{\ket{\psi_I}\}$ through, e.g., random state preparation or VQE methods applied to a chosen LGT Hamiltonian.
        
        \item Use the Pauli decomposition \eqref{bare_op_decomp} to identify the support $\Supp O_m^{(0)}$ via \eqref{pauli_support}.
        
        \item For each $P_\alpha\in\Supp O_m^{(0)}$, determine a family of orbit-supported fixed points $\{O_{\ell,\alpha}^*\}_{\ell\in L}$ as follows:
        \begin{enumerate}
            \item Choose a truncation order $n_0 \leq |\P_{3\delta q}|$ and construct the corresponding orbit \eqref{n_0-orbit}.
            
            \item Form the orbit-restricted \gls{RG} matrix $A_\alpha$ using \eqref{orb_RG_mat}.
            
            \item For each $\ell\in L$, consider an orbit-supported ansatz\footnote{Here we assume that the coefficient $\theta_0$ corresponds to the seed string $P_\alpha$. The additional constraint $\theta_0 = 1$ is imposed to remove the overall scaling ambiguity discussed in \Cref{app:alg_supp-orb}.},
            \begin{align}
                \label{orbit_ansatz}
                O_{\ell,\alpha}(\theta) = \sum_{P_i\in\Orb_\A^{n_0}P_\alpha}\theta_i\,P_i, && \theta_0 \coloneqq 1,
            \end{align}
            and define the residual operator,
            \begin{align}
                \R_{\ell,\alpha}(\theta) \coloneqq O_{\ell,\alpha}(\theta) - 2^{-\ell} O_{\ell,\alpha}\left(A_\alpha\theta\right).
            \end{align}
            
            \item Fix a tolerance $\varepsilon \ll 1$ and minimize the cost function \eqref{cost_fcn} over the remaining coefficients $\{\theta_i\}_{i > 0}$.
            \begin{itemize}
                \item If $\theta^*$ satisfying $\mathcal{C}(\theta^*) < \varepsilon$ is identified, set 
                \begin{align}
                    O_{\ell,\alpha}^* \coloneqq M^{-\ell}\,O_{\ell,\alpha}(\theta^*),
                \end{align}
                and proceed to the next orbit.
            \end{itemize}


            \item If no acceptable solution is obtained, restart from step (b) and look for a $\lambda=2^\ell$-eigenvector using the order-$(n_0+1)$ orbit \eqref{(n_0+1)-orbit}. 
            \begin{itemize}
                \item If a non-trivial solution is not identified once \eqref{orbit_grade} is satisfied (i.e., the grade $n_\alpha$ has been reached), set
                \begin{align}
                    O_{\ell,\alpha}^* \coloneqq 0.
                \end{align}
            \end{itemize}

        \end{enumerate}
        
        \item Reconstruct the fixed point associated with $O_m^{(0)}$ by combining all orbit-supported solutions:
        \begin{align}
            O_m^* \coloneqq \sum_{\ell\in L}\left(\sum_{P_\alpha\in\Supp O_m^{(0)}} c_{\ell,\alpha}^{(0)}\,O_{\ell,\alpha}^*\right).
        \end{align}
    \end{enumerate}
\end{alg}

Conceptually, VAPOR may be viewed as a variational Krylov-subspace method adapted to \gls{RG} flows of local Pauli operators. The role of the Krylov construction is twofold: first, it identifies the minimal \gls{RG}-invariant sectors relevant to the flow of each Pauli string, and second, it provides a systematically improvable hierarchy of variational search spaces through the nested sequence
\begin{align}
    \K_{n_0}(\A,P_\alpha)
    \subset
    \K_{n_0+1}(\A,P_\alpha)
    \subset \cdots \subset
    \K_{n_\alpha}(\A,P_\alpha).
\end{align}
In this sense, the orbit expansion procedure plays a role analogous to adaptive Krylov-subspace enlargement in conventional iterative eigensolvers \cite{Saad2011}, but formulated directly at the level of \gls{RG} flows for local Pauli operators.

\subsection{Application to the 4-Flux Operator\label{app:alg_4-flux}}
We illustrate the main concepts in the case of the 4-flux operator \eqref{4-flux_loc} and the filling kernel \eqref{local_embed}. In particular, we first derive the explicit Pauli-level \gls{RG} map, then analyze the resulting Pauli orbits and orbit-supported fixed points.

Recall from \Cref{app:alg_pauli-cg} that the local embedding of computational basis states \eqref{I_fill_loc} induces the coarse-graining map \eqref{A_fill_loc} which factorizes independently across the $j,\pi,k$-label sectors. Taking an arbitrary $6q$-qubit Pauli string,
\begin{align}
    P = \left(P_{j_L}\otimes P_{\pi_L}\otimes P_{k_L}\right)\otimes\left(P_{j_R}\otimes P_{\pi_R} \otimes P_{k_R}\right),
\end{align}
we find that $\A^\text{fill}_{6q\to 3q}(P) = 0$ unless the following criteria are met:
\begin{enumerate}
    \item[(i)] $P_{\pi_R}\in\{I, Z\}^{\otimes q}$ (diagonal),
    \item[(ii)] $P_{j_L}\otimes P_{j_R}\in\{I,Z\}^{\otimes 2q}$ (both factors diagonal) or $P_{j_L}\otimes P_{j_R}\in\{X,Y\}^{\otimes 2q}$ (both factors off-diagonal),
    \item[(iii)] Condition (ii) is also satisfied by $P_{k_L}\otimes P_{k_R}$.
\end{enumerate}
Whenever these conditions are satisfied, we find
\begin{align}
    \label{A_fill_action}
    \A^\text{fill}_{6q \to 3q}(P) &= \A^\text{fill}_{2q \to q}\left(P_{j_L}\otimes P_{j_R}\right) \otimes P_{\pi_L} \nonumber \\
    &\qquad \otimes \A^\text{fill}_{2q \to q}\left(P_{k_L}\otimes P_{k_R}\right),
\end{align}
where $\A^\text{fill}_{2q \to q}$ acts identically on the $j,k$-label sectors, as one might expect given the form of the embedding \eqref{I_fill_loc}. Reducing to the single-qubit case, we obtain the explicit action of $\A^\text{fill}_{2\to 1}$ on $\P_1\otimes \P_1$ given in \Cref{tab:q=1_fill}. This extends naturally to an action of $\A^\text{fill}_{2q \to q}$ on $\P_q\otimes \P_q$ through the bitwise-product structure \eqref{A_fill_label}.

\begin{table}[ht]
    \renewcommand{\arraystretch}{1.5} 
    \setlength{\tabcolsep}{12pt} 
    \begin{tabular}{| c | c | c |}
        \hline
        $P_a$ & $P_b$ & $\A^\text{fill}_{2 \to 1}(P_a \otimes P_b)$ \\
        \hline
        $I$ & $I$ & $I$ \\
        $I$ & $Z$ & $Z$ \\
        $Z$ & $I$ & $Z$ \\
        $Z$ & $Z$ & $I$ \\
        \hline
        $X$ & $X$ & $X$ \\
        $X$ & $Y$ & $Y$ \\
        $Y$ & $X$ & $Y$ \\
        $Y$ & $Y$ & $-X$ \\
        \hline
    \end{tabular}
    \caption{Action of the local coarse-graining map $\A^\text{fill}_{2 \to 1}$ on products of single-qubit Pauli strings $P_a, P_b \in \P_1$, both associated with either the $j$- or $k$-label sectors. The diagonal and off-diagonal Pauli strings are separated by a horizontal line, reflecting the absence of mixing between these regimes throughout the \gls{RG} flow.}
    \label{tab:q=1_fill}
\end{table}

Turning to the 4-flux operator \eqref{4-flux_loc}, we see that the relevant local Hilbert space is $\H^\text{loc}_m = \H_{m-1}\otimes \H_m$, corresponding to $\delta_0 = 1$ and $\delta_1 = 0$. It follows that $\O_{\rho,m}^{(0)}$ admits a Pauli decomposition of the form
\begin{align}
    \O_{\rho,m}^{(0)} = \sum_\alpha c_{\alpha}^{(0)}\,P_\alpha\in\Span\P_{6q},
\end{align}
with resolution-independent coefficients $\{c_\alpha^{(0)}\}$. Under local refinements, the transformation \eqref{pauli_refine} yields
\begin{align}
    P_\alpha\xmapsto{\;6q\to 12q\;} I^{\otimes 3q} \otimes \left(P_\alpha\otimes I^{\otimes 3q} + I^{\otimes 3q}\otimes P_\alpha\right).
\end{align}
Assuming that this refined string survives coarse-graining via $\A^\text{fill}_{12 q \to 6q}$, we obtain the Pauli-level \gls{RG} map 
\begin{align}
    \label{fill_RG_map}
    \A_\text{fill}(P_\alpha) \coloneqq P_\alpha + I^{\otimes 3q}\otimes\A^\text{fill}_{6q \to 3q}(P_\alpha),
\end{align}
where the second term is evaluated using \eqref{A_fill_action}.

Because $\O_{\rho,m}^{(0)}$ acts non-trivially only on the labels $(k_{m-1},j_m,\pi_m,k_m)$, any $P_\alpha\in\Supp\O_{\rho,m}^{(0)}$ is necessarily of the form
\begin{align}
    P_\alpha = I^{\otimes q}\otimes I^{\otimes q}\otimes Q_\alpha, && Q_\alpha\in\P_{4q}.
\end{align}
Accordingly, \eqref{fill_RG_map} restricts consistently to a map $\Span\P_{4q}\to\Span\P_{4q}$, given by
\begin{align}
    \A_\text{fill}\big|_{\Span\P_{4q}}\left(Q_\alpha\right) = Q_\alpha + I^{\otimes q}\otimes \A^\text{fill}_{6q \to 3q}\left(I^{\otimes 2q}\otimes Q_\alpha\right).
\end{align}
This reflects the fact that the \gls{RG} behaviour of $\O_{\rho,m}^{(0)}$ is fully determined by the local subgraph associated with $(k_{m-1},j_m,\pi_m,k_m)$, as mentioned in \Cref{sec:test_4-flux}. However, this additional reduction will not be necessary for the remainder of this discussion.

As noted in \Cref{app:alg_supp-orb}, the Pauli support of $\O_{\rho,m}^{(0)}$ satisfies 
\begin{align}
    \big|\Supp\O_{\rho,m}^{(0)}\big| \leq 4^{2q} = |\P_{6q}|^{1/3},
\end{align}
with the precise support size depending on $\rho$. Using the \gls{RG} map \eqref{fill_RG_map}, we find that the non-trivial orbits generated from $P_\alpha\in\Supp\O_{\rho,m}^{(0)}$ fall into two distinct classes:
\begin{enumerate}
    \item[(i)] $\Orb_{\A_\text{fill}}P_\alpha = \{P_\alpha\}$:
    \begin{align}
        \A_\text{fill}\left(P_\alpha\right) = 2^\ell P_\alpha, && \ell\in\{0,1\}.
    \end{align}
    \item[(ii)] $\Orb_{\A_{\text{fill}}}P_\alpha = \{P_\alpha,P_\alpha'\}$ for some $P_\alpha'\neq P_\alpha$:
    \begin{align}
        \A_\text{fill}\left(P_\alpha\right) = P_\alpha + P_\alpha', && \A_\text{fill}\left(P_\alpha'\right) = 2\,P_\alpha'.
    \end{align}
\end{enumerate}
Note that the latter situation directly mirrors the operator-level behaviour encountered in \eqref{4-flux_filling}--\eqref{lambda_step}. In each case, the restricted \gls{RG} matrices are given by
\begin{align}\label{fill_orbit_mats}
    \text{(i)}\quad A_\alpha = 
         2^\ell, && \text{(ii)}\quad A_\alpha = \left(\begin{array}{cc}
        1 & 0 \\
        1 & 2
    \end{array}\right),
\end{align}
with the former clearly corresponding to simple scalar multiplication.

Since $\O_{\rho,m}^{(0)}$ is resolution-independent, the orbit-supported fixed points $\O_\alpha^*$ do not contain inverse powers of $M$ and must therefore correspond to strict (i.e., $\lambda = 1$) eigenvectors of the associated \gls{RG} matrices. Using the normalization constraint mentioned in \eqref{orbit_ansatz}, the orbit-supported fixed points in case (i) are given by
\begin{align}
    \text{(i)}\implies  \O_\alpha^* = \begin{cases}
        P_\alpha, & \ell = 0, \\
        0, & \ell = 1,
    \end{cases}
\end{align}
while the 2-dimensional orbits yield
\begin{align}
    \text{(ii)} \implies \O_\alpha^* = P_\alpha - P_\alpha'.
\end{align}
This latter case should once again be compared with the form of the operator-level result \eqref{4-flux_FP}. Finally, one combines all such orbit-supported contributions to obtain
\begin{align}
    \O_{\rho,m}^* = \sum_{P_\alpha\in\Supp\O_{\rho,m}^{(0)}} c_\alpha^{(0)}\,\O_\alpha^*,
\end{align}
which, upon extension to the full Hilbert space, precisely reproduces the analytical result of \Cref{app:RG}.

\section{Background on variational Quantum Computations\label{app:VQE}}

To determine the ground state properties of quantum systems with prohibitively large Hilbert spaces, a class of hybrid quantum-classical algorithms known as Variational Quantum Eigensolvers (VQEs) has emerged \cite{Cerezo:2020jpv}. In principle, the ground state could be found by directly diagonalizing the Hamiltonian and measuring its eigenvalues. However, this relies on basis-changing unitaries that require prohibitively deep circuits, making them infeasible on noisy intermediate-scale quantum (NISQ) computers. VQEs instead employ an optimization-based approach that keeps circuits shallow, accounting for the limited qubit connectivity and coherence times of current devices.

Variational algorithms define a cost function $C$ and an ansatz depending on a set of parameters $\theta$. The ansatz is then trained in a hybrid quantum-classical loop to solve the optimization problem
\begin{equation}
\theta^* = \arg \min_{\theta} C(\theta)
\end{equation}
Here, $C(\theta)$ is estimated using a quantum computer, while the parameters $\theta$ are updated via classical optimization routines. The VQE is one of the most widely used algorithms for ground state preparation and energy estimation \cite{Peruzzo2014, Fedorov:2021bcq}. Its goal is to approximate the ground state of a given Hamiltonian $H$ by minimizing the energy expectation value. Accordingly, the cost function corresponds to the energy of the trial state, defined as
\begin{equation}
E(\theta) = \bra{\psi(\theta)}H\ket{\psi(\theta)}
\end{equation}
where the trial quantum state $\ket{\psi(\theta)}$ is prepared using a quantum circuit $U(\theta)$ and an initial state $\ket{\psi_0}$
\begin{equation}
    \ket{\psi(\theta)} = U(\theta) \ket{\psi_0}.
\end{equation}

In general, the choice of the ansatz depends on the problem at hand, with several different proposals existing \cite{Bian2018QuantumCM, Gard2019EfficientSS, Ryabinkin2018QubitCC, Taube2006NewPO, Peruzzo2014, Farhi:2014ych}. Here, the variational circuit is constructed using a hardware-efficient ansatz (HEA) \cite{Kandala:2017vok}.

Because we cannot implement the deep unitary circuits required for direct diagonalization, we must find an alternative, NISQ-friendly way to evaluate the expectation value of the Hamiltonian. To achieve this, the Hamiltonian is instead decomposed into a linear combination of tensor products of Pauli operators $P_\alpha$,
\begin{equation}
    H = \sum_\alpha c_\alpha P_\alpha, \quad P_\alpha \in \{I, X, Y, Z\}^{\otimes n}.
\end{equation}
By decomposing the Hamiltonian, we reduce the complex global measurement into a sum of simpler, local measurements. Because current quantum hardware natively measures only in the computational ($Z$) basis, the expectation value of each Pauli string is evaluated individually. This is achieved by applying local single-qubit rotations that map the eigenbasis of each $P_i$ into the computational basis. By repeatedly preparing the trial state and measuring the qubits, we statistically estimate the expectation value $\langle P_i \rangle$ for each string. The total energy $E(\theta)$ is then reconstructed on a classical computer via the weighted sum. Once the energy is evaluated, a classical optimizer iteratively updates the parameters $\theta$ to minimize this cost function. For our simulations, we employ the Powell optimization method initialized with random parameter candidates.

To operate within the coherence time limits of NISQ hardware, it is necessary to employ an ansatz with an inherently shallow circuit depth. The Hardware-Efficient ansatz (HEA) achieves this by abandoning problem-specific unitaries in favor of a layered sequence of native hardware operations. The number of applied layers acts as a tunable parameter, allowing us to find the optimal trade-off between algorithmic expressivity and gate overhead. Consequently, HEAs provide a highly practical route for ground state preparation on near-term quantum devices \cite{Kandala:2017vok, Cerezo:2020jpv}. 

In our implementation, the HEA prepares the trial state through a parameterized unitary operator $U(\theta)$ acting on the initial zero state. To explicitly model the closed geometry of the physical system, the ansatz is structured as a sequence of $L$ repeated layers,
\begin{equation}
U(\theta) = \prod_{l=1}^L U_{\text{ent}} U_{\text{rot}}(\theta^{(l)})
\end{equation}
where $\boldsymbol{\theta}^{(l)}$ denotes the subset of variational parameters for the $l$-th layer. The rotation block consists of independent, parameterized Pauli-$Y$ rotations applied simultaneously to each of the $N$ qubits:
\begin{equation}
U_{\text{rot}}(\theta^{(l)}) = \bigotimes_{i=0}^{N-1} R_y(\theta_i^{(l)})
\end{equation}
To generate entanglement while rigorously enforcing the periodic boundary conditions of the lattice, the entangling block $U_{\text{ent}}$ is constructed with a closed ring topology. Specifically, it consists of CNOT gates applied sequentially to adjacent qubits across the entire register:
\begin{equation}
U_{\text{ent}} = \prod_{i=0}^{N-1} \text{CX}{i, (i+1) \bmod N}
\end{equation}
where $\text{CX}{c,t}$ denotes a CNOT gate with control qubit $c$ and target qubit $t$. The modulo operator explicitly ensures that the sequence terminates with a final entangling gate connecting the $N$-th qubit back to the first, thereby closing the periodic loop at every layer.

Within our framework, described in \Cref{App:Toy_model}, the lowest cutoff of $j_\text{max}=1/2$ restricts each momentum to two possible values. This allows for a single-qubit encoding, yielding a 6-qubit system where exploring deep circuits and numerous parameter initializations is computationally straightforward. However, increasing the cutoff to $j_\text{max}=3/2$ expands the allowed values per momentum to four, doubling the system to 12 qubits. While this size remains readily simulable, it serves as a practical testbed to expose and analyze the general scaling limitations of hardware-efficient architectures. Because the HEA is problem-agnostic, it does not naturally confine its search to the physically relevant symmetry sectors of the Hamiltonian. Consequently, as the system scales, capturing the necessary entanglement requires a proportionally deeper circuit. This leads to a rapid proliferation of variational parameters, ultimately straining the classical optimization process. To mitigate this without incurring massive gate overhead, we systematically analyzed the trade-off between depth and accuracy in the 12-qubit model. Our results demonstrate that, provided an adequate ensemble of initial starting candidates, moderately shallow circuits are sufficient to successfully converge on highly accurate ground-state energies for our system.

For illustration, we present the analysis of the cutoff $j_\text{max}=1/2$ in \Cref{fig:H_circuits}. In the left panel, the analytically computed ground-state energies of the Hamiltonian are plotted as a function of the coupling constant $\lambda$. They are compared against the simulated VQE results obtained across various circuit depths and initialization trials. We observe that even with a shallow ansatz, the simulation accuracy is good, becoming excellent with just a slight increase in depth. However, because our optimization relies on random parameter initialization, the algorithm remains susceptible to starting far from the global minimum, getting trapped in local minima \cite{Cerezo:2020jpv}. Thus, a sufficient number of random trials is required to guarantee convergence. In the right panel, we display the state infidelity, defined as $1-F$, where $F$ represents the fidelity. This metric strictly quantifies how closely the variationally generated trial state approximates the true analytical ground state. While the ideal value is zero, the data indicates that as $\lambda$ increases, i.e., as the magnetic Hamiltonian becomes more dominant, achieving high accuracy with shallow circuits becomes progressively more difficult. This growth in error suggests that a stronger coupling drives the system into a highly entangled regime, which the limited depth of a standard HEA struggles to fully span. Nevertheless, mirroring the energy trends, deploying slightly deeper circuits yields results much closer to the ideal value.
\begin{widetext}
    \begin{minipage}{0.9\linewidth}
        \centering
        \includegraphics[width=\textwidth]{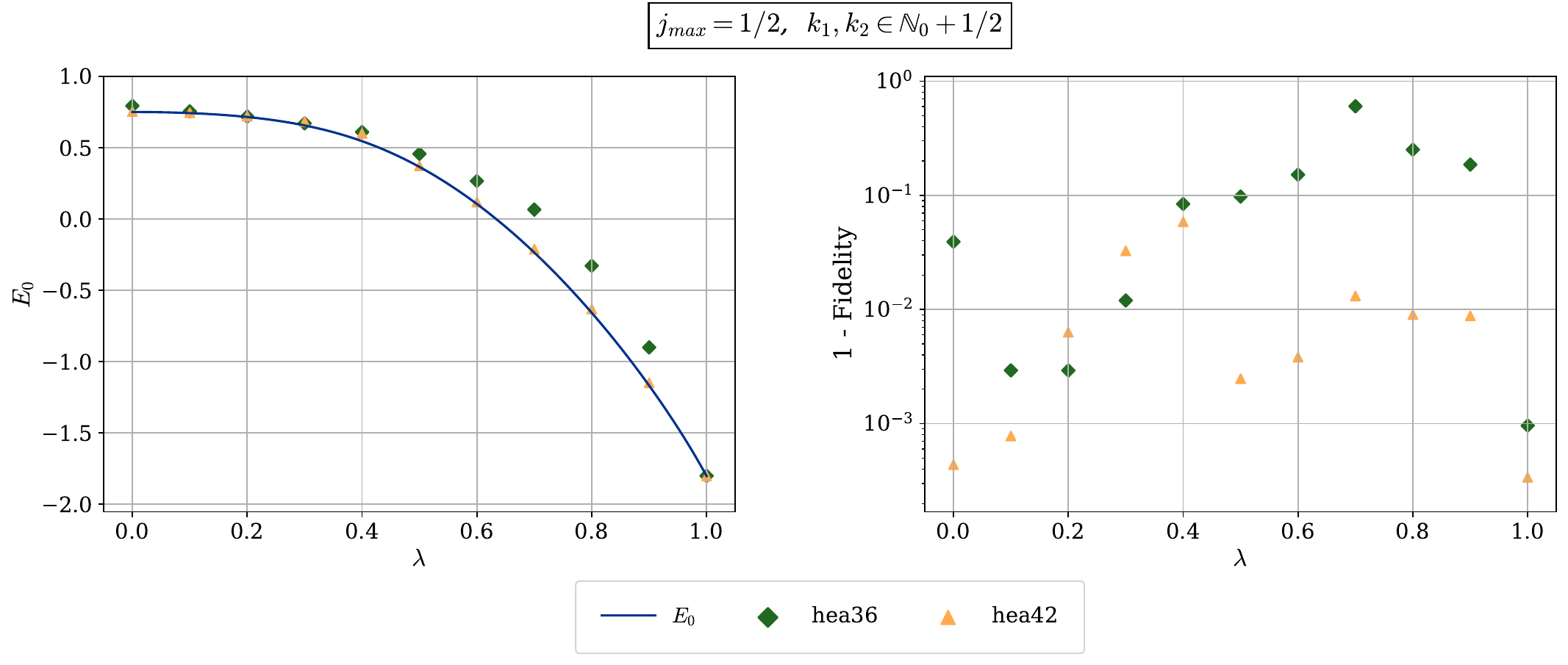}
        \label{vqe-half0.5}
    \captionof{figure}{VQE ground state simulation results using a HEA for $j_\text{max} = 1/2$. The left panel displays the exact analytical energies alongside the VQE-optimized energies across different coupling constants $x$ and circuit depths. The right panel shows the state infidelity ($1 - F$), on logarithmic scales. The error metric demonstrates an improvement in ansatz performance even at relatively shallow circuit depth.}
    \label{fig:H_circuits}
    \end{minipage}
\end{widetext}

\end{document}